\documentclass[12pt]{pazha2}
\usepackage{placeins}
\usepackage{ulem}
\usepackage{graphicx}
\usepackage{multirow}

\usepackage{caption}
\usepackage{color}
\definecolor{darkblue}{rgb}{0,0,0.9}

\usepackage{rotating}

\usepackage{amssymb}
\usepackage{longtable}

\usepackage{textcomp}
\def\aa{$^{\mbox{\footnotesize a}}$}
\def\bb{$^{\mbox{\footnotesize b}}$}
\def\cc{$^{\mbox{\footnotesize c}}$}
\def\dd{$^{\mbox{\footnotesize d}}$}
\def\ee{$^{\mbox{\footnotesize e}}$}

\DeclareTextFontCommand{\textbfit}{%
  \fontseries\bfdefault 
  \itshape
}

\begin{document}
\journalinfo{to be}{2024}{50}{12}{756}{783}{813}[779]
\sloppypar

\title{\bf Search for Astrophysical Transients on Limiting Time
  Scales and Their Classification Based on {\textbfit
    {INTEGRAL}\/} Data}
\year=2024

\author{G.Yu.\,Mozgunov\address{1},
A.S.\,Pozanenko\address{1,2}\email{apozanen@iki.rssi.ru},
P.Yu.\,Minaev\address{1},
I.V.\,Chelovekov\address{1},
S.A.\,Grebenev\address{1},\\
A.G.\,Demin\address{3},
A.V.\,Ridnaya\address{3},
D.S.\,Svinkin\address{3},
Yu.R.\,Temiraev\address{4},
D.D.\,Frederiks\address{3}\\
\addresstext{1}{Space Research Institute, Russian Academy of
  Sciences, Profsoyuznaya ul. 84/32, Moscow, 117997 Russia}
\addresstext{2}{National Research University ``Higher School of
  Economics'', Moscow, 101000 Russia}
\addresstext{3}{Ioffe Institute, Russian
  Academy of Sciences, St. Petersburg, 194021 Russia} 
\addresstext{4}{GlowByte Company, Moscow, 105064 Russia}
}

\shortauthor{MOZGUNOV et al.}
\shorttitle{SEARCH FOR LONG ASTROPHYSICAL TRANSIENTS}
  
\submitted{December 12, 2024}
\revised{December 20, 2024}
\accepted{December 20, 2024}

\begin{abstract}
\noindent\footnotesize
\baselineskip 10pt
We have searched for ultra-long ($\ga 100$ s) gamma-ray
transients in the data from the anticoincidence shield (ACS) of
the SPI gamma-ray spectrometer onboard the INTEGRAL orbital
observatory and classified them by machine learning methods. We
have found about 4364 candidates for such events in the SPI-ACS
data by the ``blind'' threshold search method. We have developed
an algorithm for automatic processing of their light curves that
distinguishes a candidate for transients on various time scales
and allows its duration and fluence to be determined. The
algorithm has been applied to calculate (and compare) the fluxes
in the light curves recorded by various INTEGRAL detectors:
IREM, SPI-ACS, SPI, ISGRI, and PICsIT. These fluxes have been
used to train the classifier based on gradient
boosting. Subsequently, we have performed a cluster analysis of
the candidates found by the dimensionality reduction and
clustering methods. In conclusion we have compared the remaining
candidates with the data from the Konus-WIND gamma-ray
detectors. Thus, we have confirmed 16 candidates for
astrophysical transients, including four candidates for
ultra-long gamma-ray bursts from the events detected by the
SPI-ACS detector. Out of the probable events, but unconfirmed by
other experiments, up to 270 events can be classified as real
gamma-ray bursts.\\

\noindent
{\bf DOI:} 10.1134/S1063773725700136\\

\keywords{\sl gamma-ray transients, cosmic gamma-ray bursts,
  solar flares, geophysical events, extended emission, machine
  learning methods, gradient boosting, clustering.}

\end{abstract}

\section{INTRODUCTION}
\noindent
There are at least two types of cosmic gamma-ray bursts (GRBs)
known.  The first type is related to the core collapse of
massive stars, as confirmed by numerous observations of type Ic
supernovae (see, e.g., Woosley 1993; Paczynski 1998; Galama et
al. 1998; Kano et al. 2017; Volnova et al. 2017; Belkin et
al. 2020, 2024) associated with long GRBs. The second type
(short bursts) was predicted from the merger of a system of two
neutron stars (Blinnikov et al. 1984; Paczynski 1986) and
confirmed by the detection of GRB~170817A and a kilonova (Abbott
et al. 2017a; Pozanenko et al. 2018) from the gravitational wave
event GW 170817 caused by a neutron star merger (Abbott et
al. 2017b) and the detection of GRB~190425A, unfortunately, only
in the gamma-ray range (Pozanenko et al. 2019) caused by
historically the second recorded gravitational wave event
GW~190425 (Abbott et al. 2020) due to a neutron star merger.

The existence of two populations was first assumed while studying
the duration distribution of GRBs detected in the Konus
experiments (Mazets et al. 1981). Subsequently, this assumption
was confirmed by studying the bimodality of the distribution of
GRBs detected in the BATSE/CGRO experiment in duration parameter
$T_{90}$ (Kouveliotou et al. 1993). Follow-up studies (see,
e.g., Tarnopolski 2016) showed that this distribution is best
fitted by the sum of a (logarithmically) normal (short bursts
corresponding to a binary neutron stars merger) and a ``skewed''
log-normal distribution (long bursts).

The skewness of the distribution of a subsample of long bursts
may be related to various selection effects (see, e.g., Minaev
and Pozanenko 2020). The event duration can be distorted, for
example, due to the eclipse of the source by the Earth or an
unstable background level in the case of low-orbit satellites,
such as Swift and Fermi, and because of the limitations on the
volume of data recording into the telemetry of space
observatories (for example, Konus-WIND). The variability time
scale of the background signal at near-Earth observatories
(Fermi and Swift) can be comparable to the duration of the GRBs
themselves, given the duration of their extended emission
(Mozgunov et al. 2021). Difficulties can also arise when
recording (when the automatic system is triggered) long dim
events (in this case, their fluence can even exceed the fluence
from typical events) --- the trigger algorithms of most
experiments are adjusted to search for sufficiently short
significant excesses of the measured flux above the background
signal.  For example, the INTEGRAL Burst Alert System (IBAS),
which automatically analyzes the data from the IBIS/ISGRI and
SPI-ACS detectors onboard the INTEGRAL observatory, operates on
time scales of only up to 5~s (Mereghetti et al. 2003). We know
cases where IBAS left even moderate-duration GRBs unnoticed
(Grebenev and Chelovekov 2007; Minaev et al. 2012; Chelovekov et
al. 2019). The duration of the pulses constituting the GRB light
curve increases with decreasing lower boundary of the detection
energy range (Fenimore et al. 1995). Accordingly, the duration
of the entire GRB also depends in a similar way on the lower
boundary of the energy range. The listed features and
distortions make only an incomplete list of causes leading to
selection effects when determining the duration and the missing
of long dim events.

On the other hand, there are many physical models that predict
the existence of long-duration GRBs (in excess of 1000 s). Among
the possible progenitors of such bursts are population III
stars, supermassive low-metallicity blue giants probably formed
in the early Universe (see, e.g., Gendre et al. 2013; Gendre
2014). Owing to the enormous mass of the progenitor star, a
massive accretion disk capable of providing a long operation of
the GRB central engine through accretion onto the black hole can
be formed during the collapse of its core. If, in addition, such
supermassive primordial stars rotated rapidly, then
supercollapsars with massive magnetically dominated jets that
manifested themselves as ultra-long hard X-ray bursts could be
formed at the end of their life (Komissarov and Barkov 2010;
Barkov 2010). It is important that in the reference frame of an
observer on Earth the light curves of such bursts will be
additionally stretched noticeably in time because of their high
cosmological redshift $z$ typical for this class of objects.
The same cosmological effect will soften the spectra of GRBs
relative to their true hardness. The observation of ultra-long
GRBs can be associated with geometrical effects --- the larger
the angle between the axis of the relativistic jet of the GRB
central engine and the direction to the observer, the longer the
GRB duration for the observer (see, e.g., Janka et
al. 2006). The extended emission explained by the additional
release of energy by the protomagnetar (Metzger et al. 2011)
formed through the collapse can also be responsible for the
atypically long burst duration. In normal conditions this model
explains durations $\sim 10-100$ s, but at special values of the
magnetic field and rotation period of the magnetar the duration
can reach $\sim25\,000$ s (Dall'Osso et al. 2011; Gendre et
al. 2013).

One way to find the missing dim long-duration events is a
``blind'' search for transients using a special triggering
algorithm adjusted to search for small signal excesses above the
background on long time scales. Such searches for astrophysical
transients, but on shorter time scales, have already been
conducted, for example, based on data from the ISGRI (Chelovekov
et al. 2006, 2019; Chelovekov and Grebenev 2011), JEM-X
(Chelovekov et al. 2017), SPI-ACS (Rau et al. 2005, Savchenko et
al. 2012), SPI (Minaev et al. 2014), and PICsIT (Rodi et
al. 2018) detectors onboard the INTEGRAL observatory. The
searches can be performed by statistical methods, for example,
by the ``moving average'' method (Savchenko et al. 2012) or using
Bayesian blocks (Scargle et al. 2013). Biltzinger et al. (2020)
demonstrated the possibility of using the physical modeling of
the background instead of its purely empirical description,
which can improve the accuracy of the background subtraction
when extracting the useful signal. The up-to-date methods imply
using neural networks, which, however, require a large volume of
high-quality data for training (see, e.g., Crupi et al. 2023;
Sadeh 2019; Parmiggiani et al. 2023). The background modeling to
search for short events is usually performed by analytical methods ---
by fitting with polynomials of various degrees. In the case of
near-Earth spacecraft, depending on the in-orbit position,
fourth- or fifth-degree polynomials can be used (Arkhangelskaja
and Arkhangelskiy 2016), whereas linear fitting is suitable for
the SPI-ACS detector onboard the INTEGRAL observatory on short
time scales (Minaev et al. 2010; Bisnovatyi-Kogan and Pozanenko
2011). The events found are classified by the method of
cross-matching with the data of other experiments, by comparing
the spectral-timing properties of the event with the values
typical for various transients, and by localizing the source in
the sky (in the case of using coded-aperture telescopes).  The
up-to-date methods suggest the classification by a machine
learning method, in particular, by the random forest (Lo et
al. 2014; Farrell et al. 2015; Yang et al. 2022) or neural
network (Sadeh 2019) algorithms. Many papers are devoted to an
overview of the results of studies and theoretical models of
cosmic GRBs (see, in particular, Levan 2018; Pozanenko et
al. 2021).

In this paper we carry out a ``blind'' search for ultra-long
transients in the SPI-ACS data. We use the synergy of
statistical modeling (to search for and process gamma-ray
transients) and machine learning (to analyze our results and to
classify the detected transients) methods.

\section{INSTRUMENTS AND DATA PROCESSING}
\subsection*{\sl 2.1. The INTEGRAL Observatory\/}

\noindent
The INTEGRAL satellite was put into a highly elliptical orbit
with an initial orbital period of 72 h and an apogee $\sim
153\,000$ km. Such an orbit provides background stability on
long time scales compared to spacecrafts in near-Earth
orbits. More than 90\% of the time the satellite is outside the
Earth's radiation belts in the region of a weak magnetic
field. As a result, the spacecraft is continuously exposed to
solar and galactic cosmic rays, which contribute significantly to
the background count rate. For this reason, the mean value of
the background increases, but its stability improves.  A
high-apogee orbit (72 h and 68 h after 2015) allows one not only
to trace the evolution of the background when passing through
the Earth's radiation belts, but also to study the large-scale
behavior of the background on time scales of more than three
days.

The main instrument being used in this paper is SPI-ACS. It is
the anticoincidence shield of the cooled germanium gamma-ray
spectrometer SPI and is the most massive detector capable of
recording GRBs that has been operated in space over the entire
history of observations. Ninety one BGO (bismuth germanate)
scintillators are used as detectors. Two photomultiplier tubes
(PMTs) are coupled to each BGO crystal; the counts from all PMTs
are recorded in a single energy channel. The lower threshold of
the channel is $\sim80$ keV; its upper threshold is $\sim10$
MeV. The SPI-ACS experiment can record photons from all
directions, but the direction coincident with the axis of the
main INTEGRAL telescopes is least sensitive (within the SPI field
of view with a radius of $16^\circ$).  The time resolution of
the SPI-ACS detector is 50~ms (von Kienlin et al. 2003).

The scintillation detectors are capable of recording charged
particles as efficiently as photons. The detection results from
the recombination of an electron knocked out of one of the
crystal atoms. The primary electron knocking-out can be caused
by both a photon and a particle. Rau et al. (2005) showed that
most of the short peaks ($<0.25$~s) in the record of the
SPI-ACS count rate originate from high-energy cosmic rays.

Apart from SPI-ACS, there are also other instruments onboard the
INTEGRAL observatory. One of them is the IREM high-energy
charged-particle detector. Its main task is to monitor the
radiation environment for the timely protection of the
electronics of the scientific instruments from intense fluxes of
charged particles. It consists of three semiconductor silicon
detectors, each with a thickness of $0.5\ \mbox{mm}$, two with
an area of $25\ \mbox{mm}^2$ and one with an area of
$50\ \mbox{mm}^2$. The time resolution is 60~s. The flux is
distributed in 15 channels that differ by the energy bands and
the response curves. In this paper we use the data of the TC3
channel, since it has the widest coverage in energy (the lower
threshold is 0.8 MeV for electrons and 10 MeV for protons). The
\mbox{JEM-X} (Lund et al. 1999), SPI (Vedrenne et al. 2003),
IBIS/ISGRI (Lebrun et al. 2003; Labanti et al. 2003; Quadrini et
al. 2003), and IBIS/PICsIT (Di Cocco et al. 2003) telescopes
that differ in field-of-view width and energy band, from the
standard X-ray one for \mbox{JEM-X} to the soft gamma-ray one
for SPI and IBIS/PICsIT, together can give a broader coverage of the
transient that fell within their field of view: the energy
spectrum, the temporal structure, and the origin (particles or
radiation).

\subsection*{\sl 2.2. Background Modeling on Various Time Scales}
\noindent
We started our analysis with estimation of the maximum
accessible time scales for the search for transients in the
archival data. The maximum timescale corresponds to the maximum
time interval on which the background can still be decribed by
polynomial models. Using Kevin Hurley's
catalog\footnote{www.ssl.berkeley.edu/ipn3/masterli.txt} and the
results of Mozgunov et al. (2021), we can distinguish the
intervals during which no transients were recorded.  These
intervals are used to calculate the functional $\chi^2$ for
various time scales.

The calculation procedure consists of the following
steps:
\begin{enumerate}
    \item A background interval with a duration of 0.5 s is
      chosen. The minimum time resolution of the SPI-ACS
      detector is 50 ms and, consequently, the interval in the
      initial time resolution contains ten bins.
    \item The flux within this interval is fitted by four
      analytical models: constant, linear, third-, and
      fifth-degree polynomials.
    \item The value of the functional $\chi^2$ is calculated.
      The value of $(F\times k)^{1/2}$, where $F$ is the flux in
      a given bin and $k$ is the ``super-poissonness''
      coefficient for a given interval, is used as a $1\sigma$
      error. It is calculated as the ratio of the variance to
      the mean value of the background.
    \item The fitting interval is expanded by a factor of 2 and
      is binned in such a way that the final number of bins is
      10. This is necessary to compare the values of the
      functionals per degree of freedom, $\chi^2/$d.o.f.,
      between themselves, since in this case the distributions
      will have the same number of degrees of freedom. Since the
      number of points is always 10, the number of degrees of
      freedom is 9, 8, 6, and 4 for different models,
      respectively. 
    \item Steps 2--4 are repeated until the background interval
      exceeds a duration of 60\,000 s.
\end{enumerate}
\begin{figure*}[t]
\centering
\begin{minipage}{0.50\textwidth}
  \centering
  \includegraphics[width=0.99\linewidth]{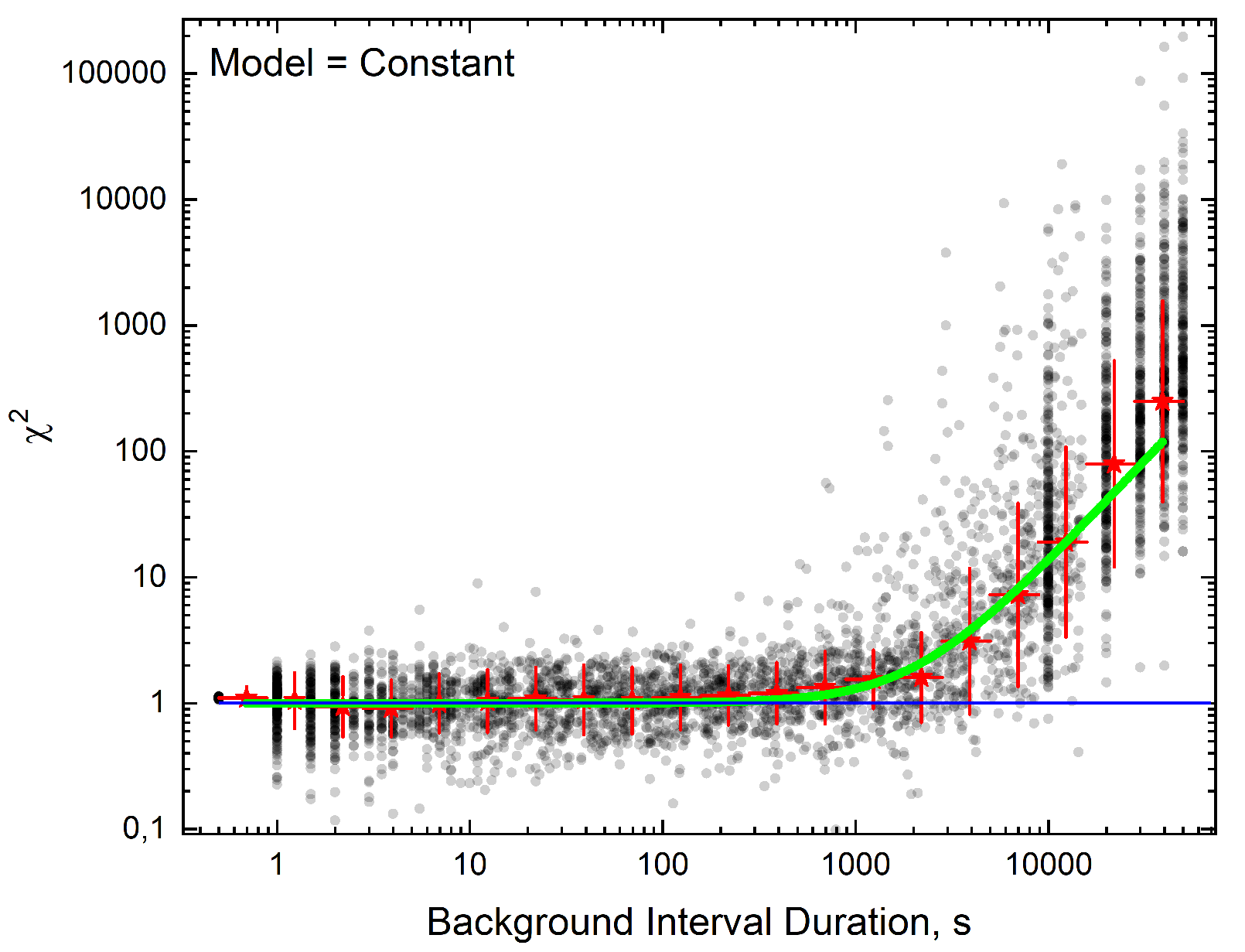}
\end{minipage}\hspace{-0.5mm}\begin{minipage}{0.50\textwidth}
  \centering
  \includegraphics[width=0.99\linewidth]{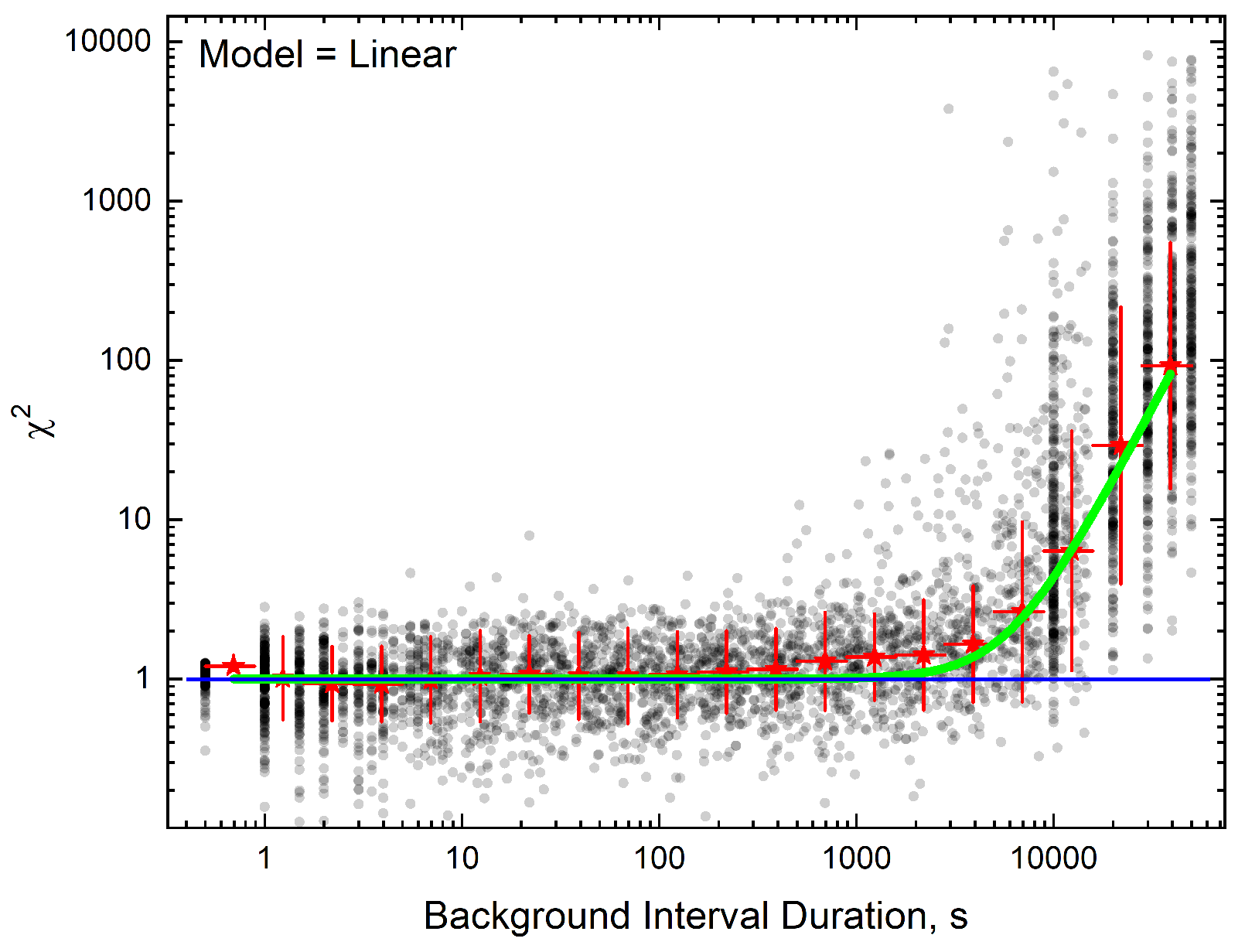}
\end{minipage}

\caption{\small\rm The result of modeling the background by
  various analytical models. The black dots indicate the results
  for the individual intervals, the mean and $1/2$ FWHM
  (half-width at half maximum) for the distribution of points
  within the corresponding duration range. The blue line marks
  the $\chi^2=1$ level, and the green line is the fit to the red
  points by the function (\ref{beuerman}). The left and right
  panels show the constant and linear model fits,
  respectively. The length of the background modeling interval
  is along the horizontal axis and the value of the functional
  $\chi^2/$d.o.f. corresponding to a given interval and model is
  along the vertical axis.}
\label{const_linear_chi2}
\end{figure*}
Steps 1--5 are repeated for $\sim180$ randomly chosen orbits
during which no bright transients were observed. Thereafter, the
range of intervals in duration from $0.5$ to $6\times10^4$ s is
broken up into 20 groups distributed uniformly in logarithmic
space. In each group we construct the $\chi^2$ distribution that
is fitted by a normal distribution. As, $\chi^2$, inherent in a
given group we use the value corresponding to the maximum in the
fit by a normal distribution and take the error as the
half-width at half maximum ($1/2$ of the FWHM). Thereafter, the
grouped values are fitted by a power law with a break,
\begin{equation}\label{beuerman}
  F(D) = A\times \left[\left(\frac{D}{D_c}\right)^{\alpha\cdot w}+
    \left(\frac{D}{D_c}\right)^{\beta\cdot w}\right]^{-1/w},
\end{equation}
in which the parameter $w=-3/2$ is fixed. Using it, we determine
the critical duration $D_c$ --- the position of the break in the
function. At background interval durations greater than $D_c$
the chosen analytical background model ceases to describe the
actual data. The modeling, grouping, and fitting results are
presented in Figs.~\ref{const_linear_chi2} (for the constant and
linear model fits) and \ref{3_5_polynom_chi2} (for the third- and
fifth-degree polynomial fits). The positions of the break are
presented in Table~\ref{critical_durs}.
\begin{figure*}[t]
\centering
\begin{minipage}{0.50\textwidth}
  \centering
  \includegraphics[width=0.99\linewidth]{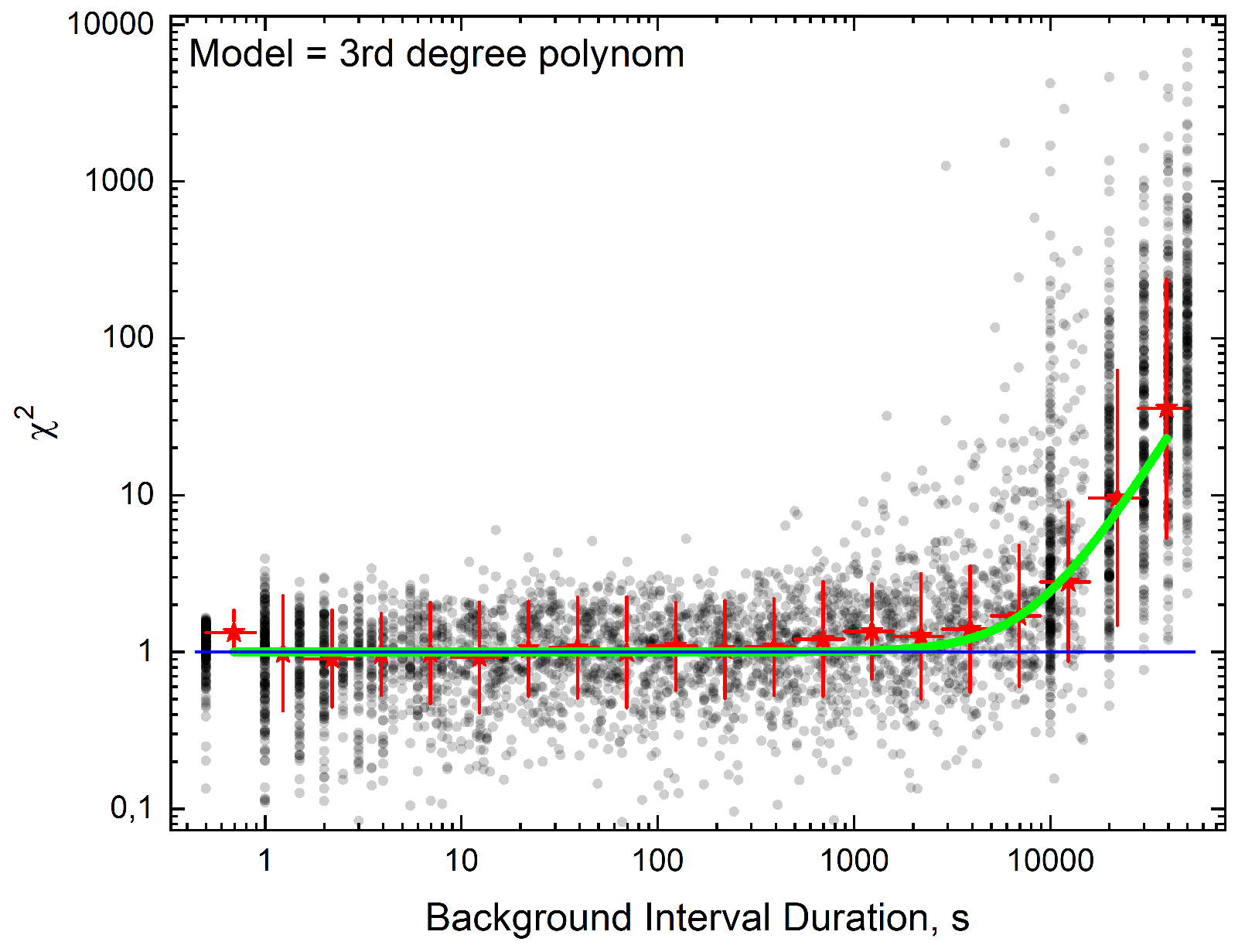}
\end{minipage}\hspace{-0.5mm}\begin{minipage}{0.50\textwidth}
  \centering
  \includegraphics[width=0.99\linewidth]{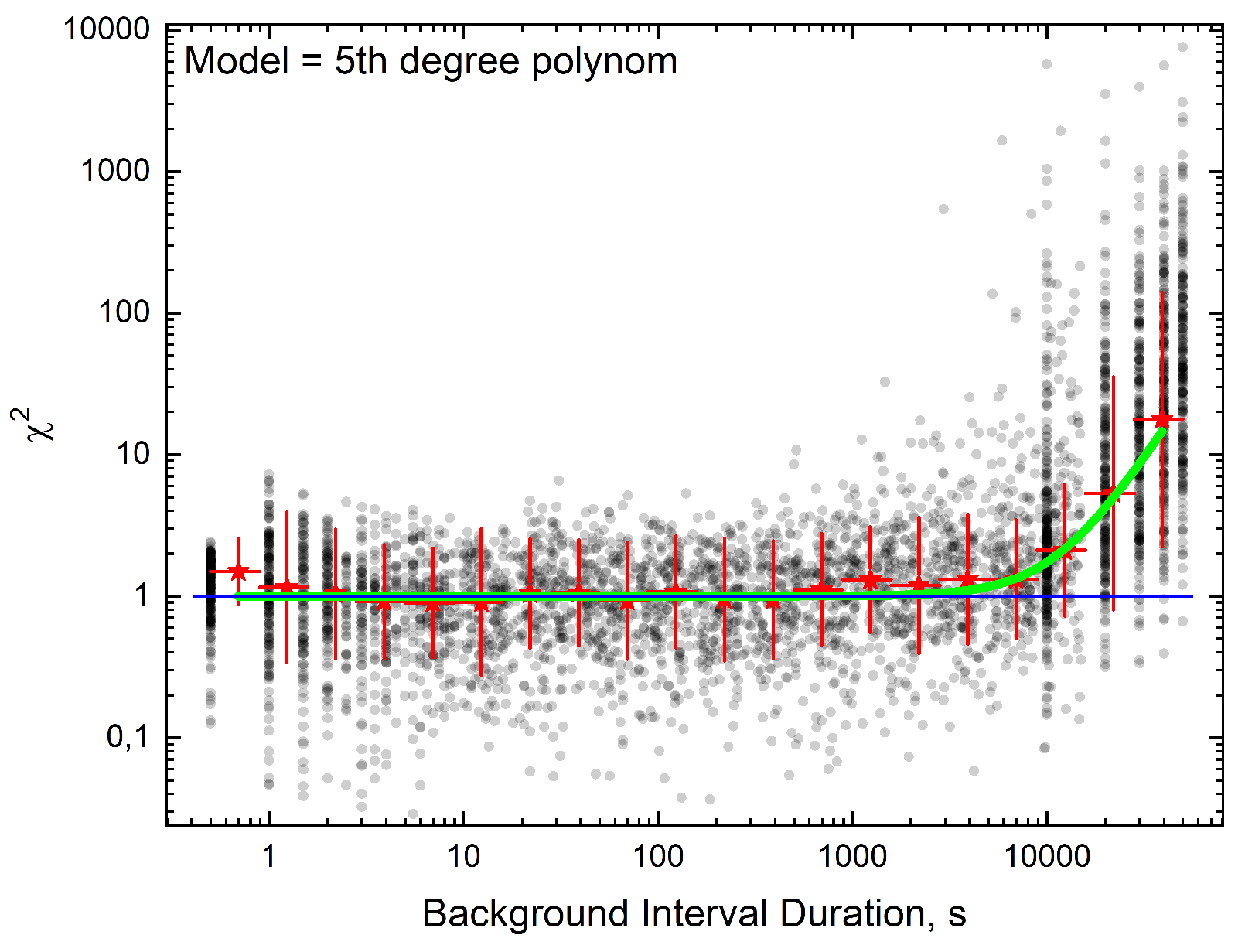}
\end{minipage}
\caption{\small\rm The result of modeling the background by
  various analytical models. The black dots indicate the results
  for the individual intervals, the mean and $1/2$ FWHM (half
  width at half maximum) for the distribution of points within
  the corresponding duration range. The blue line marks the
  $\chi^2=1$, and the green line is the fit to the red points by
  the function (\ref{beuerman}). The left and right panels show
  the third- and fifth-degree polynomial fits, respectively. The
  length of the background modeling interval is along the
  horizontal axis and the value of the functional
  $\chi^2$/d.o.f. corresponding to a given interval and model is
  along the vertical axis.}
\label{3_5_polynom_chi2}
\end{figure*}
\begin{table}[ht]
\caption{Positions of the break $D_c$ for various models}\label{critical_durs}

\centering

\begin{tabular}{l|c}
\multicolumn{2}{c}{}\\ [-3mm]
\hline
& \\ [-4mm]
\multicolumn{1}{c}{Background model}& $D_c,$ s \\
\hline
Constant & $2080 \pm 230$ \\
Linear & $5940 \pm 580$ \\
Third-degree polynomial~~~~ & $8330 \pm 910$ \\
Fifth-degree polynomial~~ & $11530 \pm 1030$ \\
\hline
\end{tabular}
\end{table}

Using the graphs in Figs.~\ref{const_linear_chi2} and
\ref{3_5_polynom_chi2} and the data from
Table~\ref{critical_durs}, we can establish the maximum duration
of the background interval fitted by simple analytical 
models. It is $\sim10^4$ s. Obviously, the maximum event
duration in such a search will be $\lesssim10^4$ s.
\begin{table*}[t]
  \caption{\small\rm Parameters of the sliding window as a
    function of the time resolution}\label{D_N} \centering
\begin{tabular}{c|c|c|c|c}
 \multicolumn{5}{c}{}\\ [-2.5mm] 
\hline
Time & Number & Window& Background & Significance\\
resolution, s& of bins $N$ & duration, s & model & threshold,
$\sigma_{T}$\\ \hline
& & & &\\ [-4mm]
1000 & 9   &  9\,000 & Linear                  & 3\\
300  & 29  &  8\,700 & Third-degree polynomial & 5\\
120  & 29  &  3\,480 & Fifth-degree polynomial & 7\\
\hline
\end{tabular}
\end{table*}

\subsection*{\sl 2.3. Search for Ultra-long Astrophysical Transients}
\noindent
We chose three main search time scales: 1000, 300, and
120 s. The SPI-ACS data spanning $\sim20$ years of observations
are formed in three light curves corresponding to these time
resolutions.  Based on the results in
Figs.~\ref{const_linear_chi2} and \ref{3_5_polynom_chi2} and
the data in Table~\ref{critical_durs}, we chose an appropriate
background model for a given time resolution and the interval
duration.  The maximum size of the interval being analyzed did
not exceed 9\,000 s. The significance thresholds were chosen so
that the number of triggers on all time scales was approximately
the same. For each time resolution we performed
``sliding-window'' processing using the following procedure:
\begin{enumerate}

\item \uwave{Data interval extraction.\/} From the light curve
  we extract the interval of $N$ successive bins in which we
  analytically model the background and search for an event. It
  is very important that there be no ``gaps'' in the data.  The
  gaps can arise, because the SPI-ACS detector may temporarily
  not transmit the data, for example, because of problems with
  telemetry.  If the difference between two successive bins is
  larger than the expected value (time resolution), then we
  assume that this is a gap in the data, and if it is found
  within the current window, then it is skipped, and the
  algorithm passes to the next window.
    
  $N$ changes, depending on the chosen time resolution; the
  correspondence between them is given in Table~\ref{D_N}. The
  window is broken up into two ranges: the event being studied
  in which the flux and the significance above the background
  are calculated. It is located in the central (number $N/2+1$)
  bin.
    
\item \uwave{Analytical background modeling.\/} The background
  model is chosen according to Table~\ref{D_N}. The linear
  background model is used for the time scale of 1\,000~s. In
  this case, the final interval duration $D$ is $9\,000$~s,
  which is greater than $D_c$ for the chosen model. This choice
  is dictated by the small number of background points (8). The
  number of degrees of freedom decreases when using a more
  complex model.  For this reason, the $\chi^2$/d.o.f. value of
  the model calculated in the background intervals decreases,
  but the systematic error introduced by the model choice
  increases.

\item \uwave{Quality-of-fit analysis.\/} The value of the
  functional $\chi^2$/d.o.f. is calculated in the background
  interval. The errors of the flux (count rate) are calculated as
  $(F\times k)^{1/2}$, where $F$ is the flux in a bin and $k$ is
  the super-poissonness coefficient determined for the
  investigated interval. This value is compared with the normal
  one for a given interval duration according to
  Figs.~\ref{const_linear_chi2} and \ref{3_5_polynom_chi2}. If
  the value of the functional is outside the $\pm 1\sigma$
  region in the corresponding group of durations, then the fit
  is recognized as unsatisfactory, and the current window is
  excluded from further consideration.

\item \uwave{Flux calculation.\/} The best background model is
  subtracted from the input data. The event flux (count rate)
  $F$ is the flux in the central bin. Its significance is
  calculated as $\sigma = F/(B_{model}\times k)^{1/2}$. If
  $\sigma<\sigma_{\rm T}$ for a given time resolution, then the
  event is excluded from further consideration.
\end{enumerate}

After steps 1--4, the time windows is shifted by 1 bin forward
along the time axis and the procedure is repeated. Thus, the
entire light curve spanning $\sim20$ years is investigated. The
event time $T_0$ is the time corresponding to the central bin in
the sliding window. As a result, we found 4364 excesses of the
count rate above the background.

\subsection*{\sl 2.4. Study of Potential Candidates for
  Astrophysical Transients on Shorter Time Scales\/}
\noindent
The generation of a list of potential candidates on the time
scales of 1000, 300, and 120 s is followed by the procedure of
their analysis: obtaining a light curve with a higher time
resolution, constructing a more accurate background model, and
determining the duration and fluxes. For this purpose, we use the
SPI-ACS data in intervals of $\pm6\,000, \pm2\,000,$ and $\pm
600$ s relative to the time $T_0$ for the events found on the
time scales of 1\,000, 300, and 120 s, respectively.

\begin{figure*}[ht]
\center{\includegraphics[width=1\linewidth]{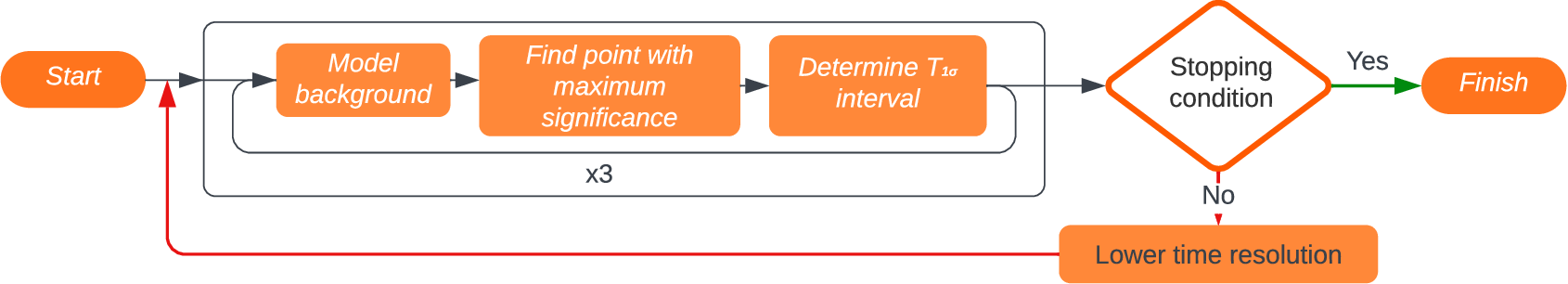}}
\caption{\small\rm The block diagram describing the transient
  processing procedure.}
\label{block_scheme}
\end{figure*}

We recursively procees the transient candidates starting from
the maximally large time resolution and gradually reduces it
until the stoping criterion is reached: either the duration was
determined with a sufficient accuracy or a limiting time
resolution of 3~s was reached. The initial time resolution
depends on the time scale on which the transient was found. For
example, it is 200 and 20 s for the event found on the 1000- and
120-s time scales, respectively. In each recursion step the
background is fitted by a third-degree polynomial by taking into
account the results from the previous step to increase the
accuracy. A block diagram of the process is shown in
Fig.~\ref{block_scheme}. The algorithm was described in more
detail by Mozgunov et al. (2024).

As a result, for each event we determined the event start time
and duration and calculated the fluxes in all instruments of the
INTEGRAL observatory after the background subtraction and the
value of the functional $\chi^2$ for the background model in
each instrument.

\subsection*{\sl 2.5. Cross-Matching}
\noindent
The cross-matching of the generated list of events with the
catalogs of known transients was used with a dual purpose: to
eliminate the confirmed events and to obtain the marking to
train the machine learning models. As a comparison catalog we
used a compilation from the following catalogs of GRBs and solar
flares and catalogs of gamma-ray event triggers: Konus-WIND
waiting-mode events\footnote{www.ioffe.ru/LEA/kw/wm/} and
Konus-WIND triggered
events\footnote{www.ioffe.ru/LEA/kw/triggers/}, IBAS SPI-ACS
bursts\footnote{www.isdc.unige.ch/integral/science/grb$\#$ACS},
Swift/BAT
bursts\footnote{swift.gsfc.nasa.gov/archive/grb\_table/}, the
Fermi GBM Burst
Catalog\footnote{heasarc.gsfc.nasa.gov/w3browse/fermi/fermigbrst.html},
and the Fermi GBM Trigger
Catalog\footnote{heasarc.gsfc.nasa.gov/w3browse/fermi/fermigtrig.html},
the masterlist of Kevin
Hurley\footnote{www.ssl.berkeley.edu/ipn3/masterli.txt}, the
RHESSI Flare
List\footnote{hesperia.gsfc.nasa.gov/hessidata/dbase/hessi\_flare\_list.txt},
and the GOES flare
list\footnote{ftp.swpc.noaa.gov/pub/warehouse/}.

Konus-WIND has conducted an almost continuous all-sky survey for
more than 30 years, completely covering the INTEGRAL operation
time. In the waiting mode Konus-WIND continuously records the
count rate of two detectors (S1 and S2 surveying the southern
and northern ecliptic hemispheres, respectively) in three energy
bands, $\sim20$--80, $\sim80$--350, and $\sim350$--1400~keV; the
time resolution of the record is 2.944~s. Given the gaps in the
data, the continuous record covers more than $\sim95$\% of the
time. Owing to the stable background on time scales up to
several days (mainly outside the periods of enhanced solar
activity), the Konus-WIND data allow transients with peak fluxes
$\gtrsim 4 \times 10^{-7}$~erg~cm$^{-2}$~s$^{-1}$ to be detected
(Ridnaia et al. 2020). The search for transient events in the
continuous Konus-WIND record was performed using a decomposition
into Bayesian blocks (Koz\-lova et al. 2019). As a result of the
search, we found GRBs, galactic transients, and solar flares,
including those missed by the Konus-WIND detector trigger
algorithm, that gave an excess above the background of more than
$4\sigma$.
\begin{figure*}[t]
\center{\includegraphics[width=1\linewidth]{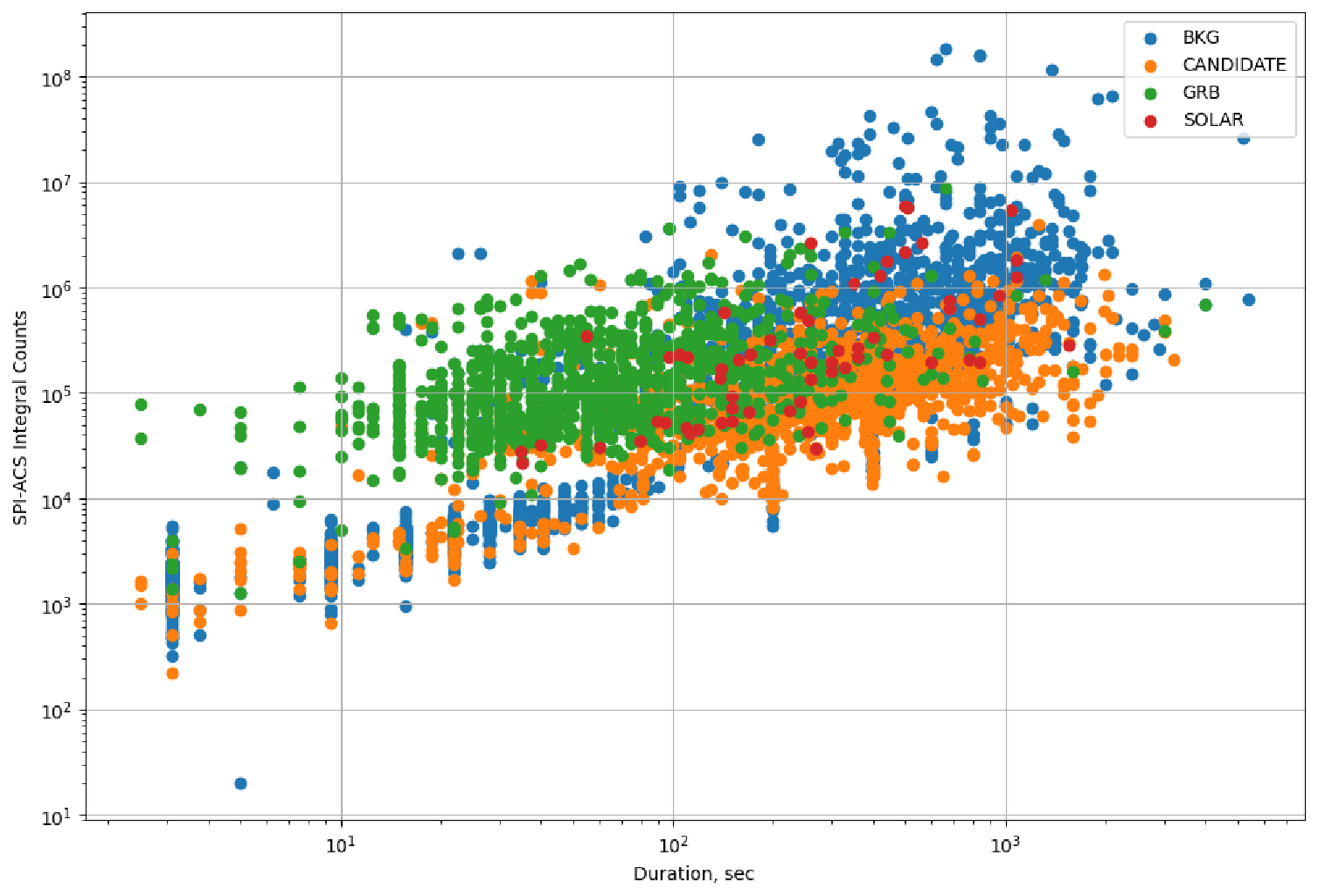}}
\caption{\small\rm The ``duration-flux'' diagram for the found
  and processed events; different colors mark four groups of
  events: solar ones, background ones, GRBs, and candidates.}
\label{duration_flux}
\end{figure*}

The cross-matching is performed using the tools of the {\tt
  pygrb\_lc}\footnote{pypi.org/project/pygrb-lc/} package,
written in {\tt Python} programming language. The algorithm is
structured as follows: for the candidate being studied with an
event start time $T_{0}$ and duration $D$ we calculate the event
from the comparison catalog closest to it. The difference in
seconds between the time from the catalog and $T_{0}$ is
calculated, and if it belongs to the interval $[-D;\ D]$, then
the events are deemed coincident.

The candidates are marked out into four groups:
\begin{enumerate}
    \item Solar flares --- a candidate was found in the GOES or
      RHESSI catalog or in any other catalog, but was marked as
      a solar event.
    \item GRBs --- a candidate was found in the catalog of GRBs.
    \item A background (geophysical) event --- the value of the
      functional $\chi^2$/d.o.f. when processing on the smallest
      time scale is higher than its nominal values (see
      Fig.~\ref{3_5_polynom_chi2}).
    \item Others --- a candidate was found in the catalog, but
      was not classified as a GRB or a solar flare (for example,
      a soft gamma repeater (SGR) flare).
\end{enumerate}
In some cases, when a transient was found in the catalogs of
solar flares and GRBs, a discrepancy arose between the RHESSI or
GOES catalogs and all of the remaining ones. For such cases we
introduced a system of priorities, $2 > 1 > 4 > 3$, justifying
this by the fact that the GOES and RHESSI energy ranges differ
noticeably from the operating energy range of the SPI-ACS
detector; therefore, finding the transient in these catalogs
could be a chance coincidence. The value of the functional
$\chi^2$ was the last to be taken into account after the
comparison with all the remaining catalogs and rules. Indeed,
the event parameters can also be reliably determined with an
unstable background. In this case, we just underestimate the
significance of the result; it is much more difficult to obtain
information about the event type. It is worth noting that we do
not have a reliable identification of background
events. Therefore, rule 3 does not guarantee that an event
belongs to this class,

In Section 3.2 we separately explain how efficient this approach
actually is for the classification of events.

The described algorithm was repeated for all events and for each
of the catalogs. Of 4364 potential candidates, 1935 events were
found at least in one catalog. The events found are presented in
Table~\ref{candidates} and Fig.~\ref{duration_flux}.
\begin{table*}[t]
\caption{\small\rm An example of the results of the operation of
  the blind transient candidate search
  algorithm\ee.} \label{candidates} \centering

\begin{tabular}{c|c|c|c|c|l}
\multicolumn{6}{l}{}\\ [-3mm]
\hline
$T_0$, &$T_{1\sigma}$\aa, &Fluence\bb, &$S/N$,\cc & Peak flux per 1\,s\dd,& Identification\\
 UTC & s &$10^3$ counts & $\sigma$& counts s$^{-1}$& \\
\hline
2003-02-12 04:04:53.978 & 21.8 & 5.04 &9.3 & 1598& RHESSI: Solar\\
2003-02-14 04:06:42.816 & 540  & 263  & 34.3 &  1915&GOES: Solar\\ 
& & & & &RHESSI: Solar\\
2003-02-14 09:52:27.816 & 194  & 101  & 20.8 &1822& ~~~~-- \\
2003-02-15 11:15:55.816 & 80   & 622  & 220.0 &30932&  K.Hurley: GRB \\
2003-02-15 15:45:03.816 & 360  & 1400 & 233.7 &9642&  RHESSI:
Solar\\ \hline
\multicolumn{6}{l}{}\\ [-3mm]
\multicolumn{6}{l}{\aa\ The duration of the continuous interval
  in each bin of which the signal}\\ 
\multicolumn{6}{l}{\ \ \ significance exceeds the
  background value by more than $1\sigma$ (Mozgunov et al. 2024).}\\
\multicolumn{6}{l}{\bb\ The fluence above the background level (the number of counts).}\\
\multicolumn{6}{l}{\cc\ The signal-to-background ratio for the transient in fluence.}\\
\multicolumn{6}{l}{\dd\ The peak flux on a time scale of 1~s.}\\
\multicolumn{6}{l}{\ee\ A full version of the table is accessible in electronic form at}\\
\multicolumn{6}{l}{\ \ \ grb.rssi.ru/INTEGRAL/GRB\_ACS\_candidates.txt.}\\
\end{tabular}
\end{table*}

\section{MACHINE LEARNING}
\noindent
The selected events confirmed by the data of other catalogs were
used as a training sample for the machine learning models.

\subsection*{\sl 3.1. Training the Classification Model}
\noindent
The event parameters determined in Section 2.5 were used as
features to train the classification model. The time scale on
which a given transient was found (1\,000, 300, or 120~s), the
minimum time resolution during its processing in Section 2.5,
the distance of the INTEGRAL satellite to the Earth obtained
from the INTEGRAL telemetry, and the light-curve shape were
added to them. Ten successive bins within the event duration
interval were responsible for the shape of the transient light
curve. The class markers were obtained in the previous section
by cross-matching. For the prediction we used only the first
three classes --- the size of the sample of ``other''
events is too small, and their search was not the direct
goal of this paper. Note that to train the model, those
solar events that were identified only in the GOES or
RHESSI catalogs were not used either because of the
mismatch between the energy ranges.

We used several standard classification models: logistic
regression, ``random forest'', and gradient boosting (Ke et
al. 2017). The latter model showed the best results. When
training the model, we selected the hyperparameters to maximize
the metric 
$$F_{\beta} = (1+\beta^2)\times \frac{precision\times
  recall}{(\beta^2\times precision) + recall},$$
where $\beta$ is selected manually, depending on the problem.  

The values of $\beta < 1$ penalize the precision more severely
than the recall, consistent with the goals of our paper --- to
make the most accurate algorithm for the sample of GRBs. In our
case, we choose $\beta = 0.5$. The details of the model training
were described by Mozgunov et al. (2024). The precision and the
recall for the training sample were $91\pm4\%$ and $73\pm6\%,$
respectively.

We applied the model for the unmarked data --- those burst
candidates for which no match was found in the catalogs: the
model marked 67 of them as GRBs. It is worth noting that the
unmarked and training samples do not belong to one
distribution. This is confirmed by the multidimensional
Kolmogorov-Smirnov test conducted with the same parameters as
those used for the training; the $p$-value, i.e., the
probability to reject the hypothesis about the samples from one
general population, is $< 10^{-40}$. This suggests that the
precision estimate obtained for the training sample can differ
greatly from that for the unmarked data.

\subsection*{\sl 3.2. Cluster Analysis}
\noindent
Machine learning can be used not only for the construction of
predictive models; using them, we can reduce the dimensionality
of the data in such a way that the parameters similar in
properties are close to one another in the resulting space. The
dimension of the output space can be any. However, the dimension
of 2 is chosen most frequently as the most convenient one for
human perception. One of the most popular UMAP algorithms
(McInnes et al., 2018) uses nonlinear transformations of the
initial features to obtain the mapping with the largest
variance.
\begin{figure*}
\centering
  \includegraphics[width=0.89\linewidth]{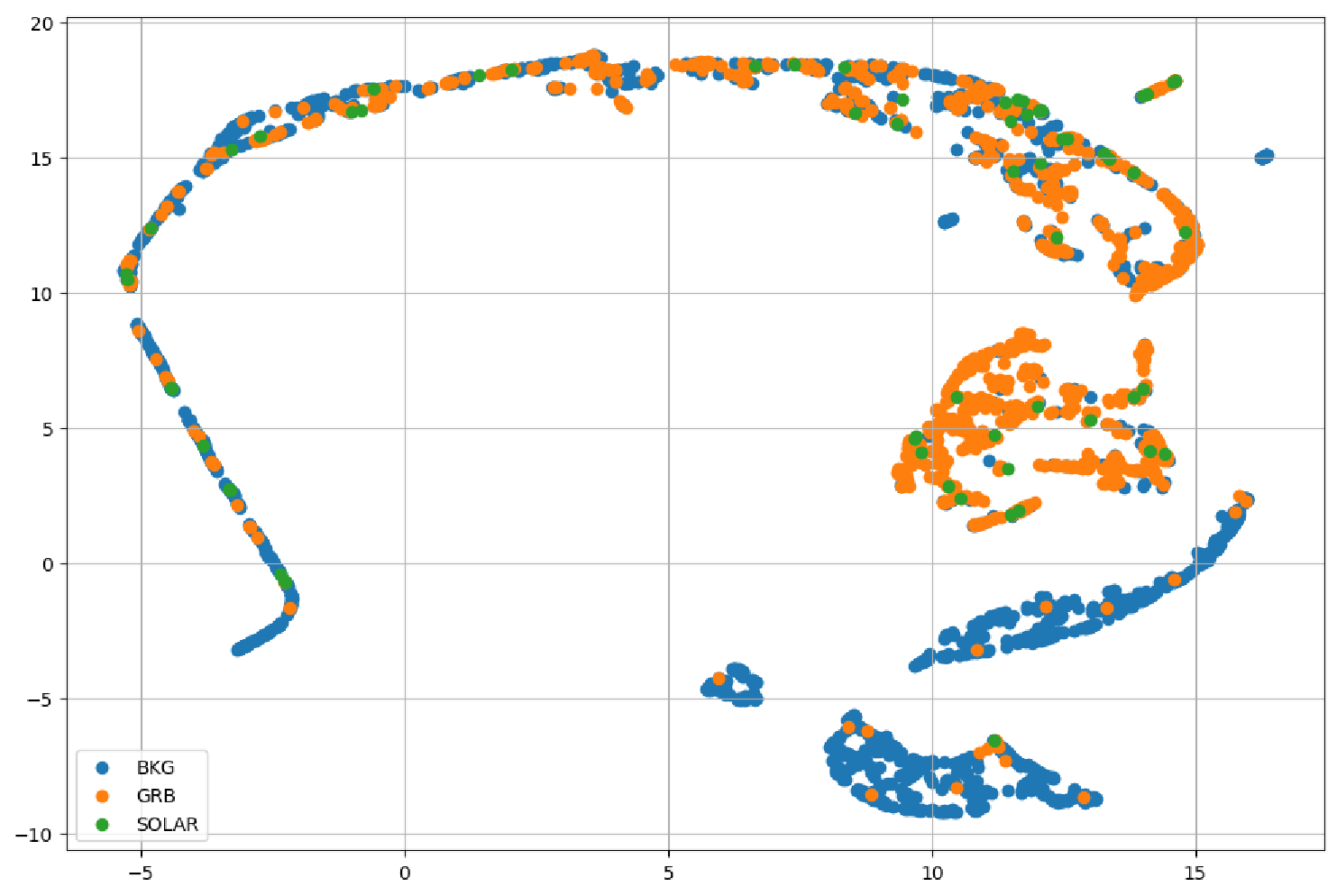}
  \includegraphics[width=0.89\linewidth]{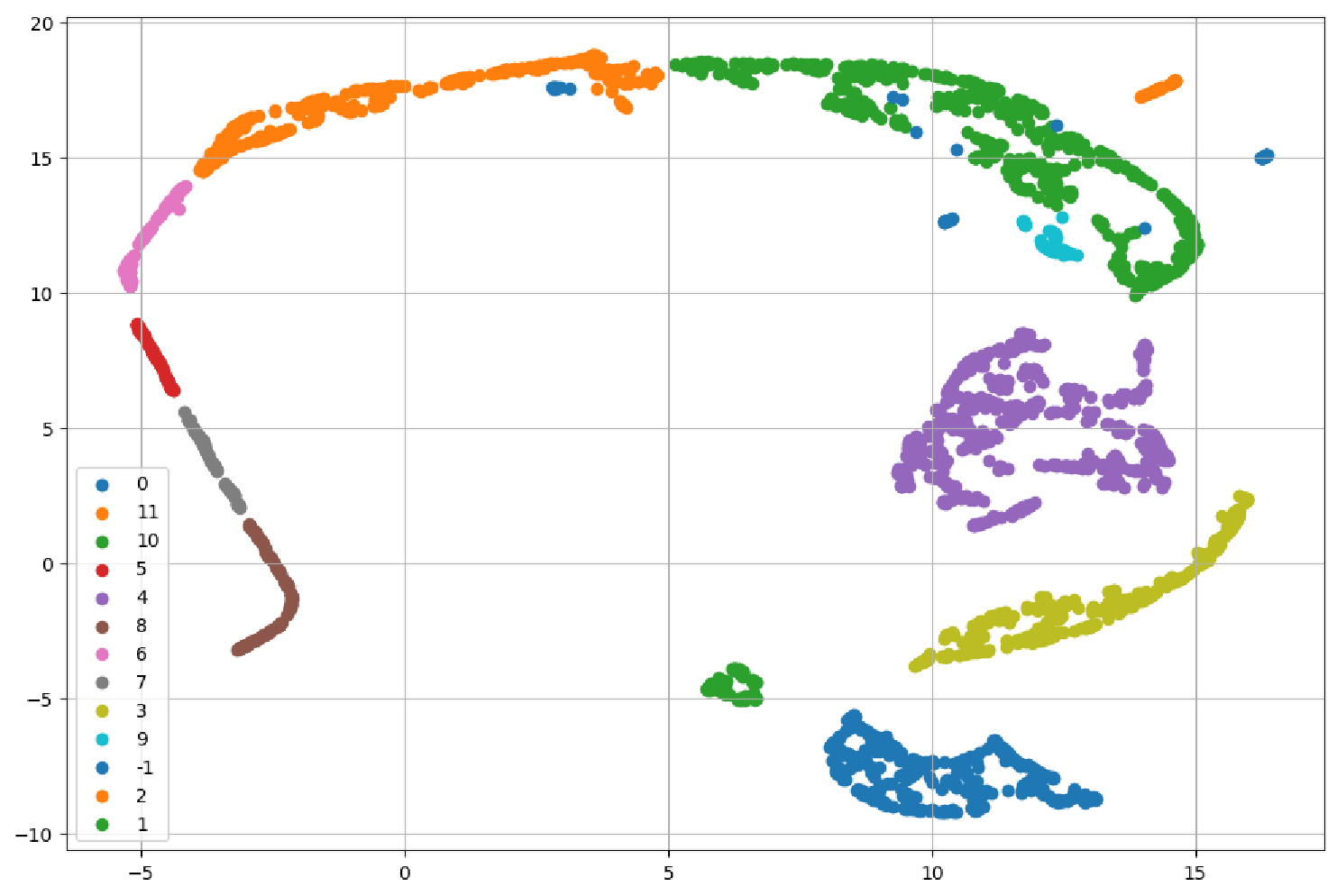}

\caption{\small\rm The result of the operation of the UMAP
  dimensionality reduction and HDBSCAN clustering
  algorithms. Abstract units are marked along the axes. On
  the upper panel the event classes are indicated by various
  colors; on the lower panel all of the clusters found are
  marked by various colors. For our analysis we used all 4364
  excesses above the background found in SPI-ACS.}
\label{clusters}
\end{figure*}

The event parameters from Section 3.1 are used as
features. Using the UMAP algorithm with standard
hyperparameters, we constructed Fig.~\ref{clusters} (upper
panel); abstract units, a nonlinear combination of initial
parameters, are marked along the axes. It can be noticed that
all events are located on one elongated curved line. This line
reflects the ``duration-flux'' correlation previously found in
Fig.~\ref{duration_flux}. There are the most energetic events in
the left part of the line and the dimmest ones in its right
part. It can also be seen that this line is nonuniform; it has
thickenings and thinnings, making the clustering possible. For
this purpose, we use the HDBSCAN algorithm. The result is
presented in Fig.~\ref{clusters} (lower panel). The cluster
numbers are physically meaningless.

For each cluster we calculated the distribution in event
types. We distinguished two clusters in which the GRB fraction
is $>50\%$; these are clusters 2 and 4. We use them to classify
the unmarked candidates.  This method marked 544 additional
events as GRBs, with this list overlapping with the classifier
results by 60 events. No separate cluster is distinguished for
solar flares. There is no difference between the two GRB clusters
in both duration and flux either. The precision of the method is
determined by the choice of clusters and the distribution of
events within them and is currently estimated to be $\sim79\%$
for the training sample.

Note that clusters 0, 1, 3, 5, 6, 7, and 8 located at the
opposite ends of the curve are completely dominated by
background events. The brightest events (at the left end of the
curve) are presumably the transients associated with charged
particles: they have a high energy in all INTEGRAL instruments;
the functional $\chi^2$ has a large value (more than 3.5),
because either part of the event falls into the background
fitting interval (because of the enormous duration) or several
events occur within one window (for example, when crossing the
radiation belts). The right end represents the random triggers
that are distinguished not through an anomalous $\chi^2$ value,
but through a low significance at the detection threshold.

\section{RESULTS}
\noindent
In this paper:\\ [2mm]
(1) We determined the limiting time scales on
which the SPI-ACS detector background could be successfully
fitted by polynomial functions; the largest time scale
$\sim10^4$~s is achieved when fitting the data by fifth-degree
polynomials. The dependence of the maximum time scale on the
degree of the polynomial can be used to test the quality of the
background fit. For this purpose, for the test background
segment we need to calculate the functional $\chi^2$/d.o.f. and
to superimpose it on the figure (Fig.~\ref{const_linear_chi2} or
\ref{3_5_polynom_chi2}) corresponding to the chosen background
model. If the value is within the range for the corresponding
duration group, then the model is suitable for describing the
chosen background segment. If not, then we should increase the
degree of the polynomial or decrease the duration of the time
interval.

(2) We conducted a blind search for long transients in the
SPI-ACS data. We found 4364 transients: 1\,325, 1\,754, and
1\,285 on the time scales of 1\,000, 300, and 120 s,
respectively. It can be seen from Fig.~\ref{duration_flux}
that the classes of the previously known events
differ: the solar flares (or their derivatives SEPE --- solar
energetic particle events) are, on average longer and
more energetic than the GRBs. Moreover, the boundary below which
there are no events is distinct --- it corresponds to the
minimum detection threshold chosen by us to search for transients.
The behavior of this dependence is described by the law
$Fluence \sim Duration^{1/2}$. In Fig.~\ref{clusters} clusters 0,
1, and 3 correspond to solar flares, the fraction of background
events (random fluctuations) in them is $\gtrsim 99\%$, and the
GRBs that fell into these clusters are most likely chance
matches with the catalogs and are actually unseen by the SPI-ACS
detector. 
\begin{figure}[t]
\center{\includegraphics[width=1.\linewidth]{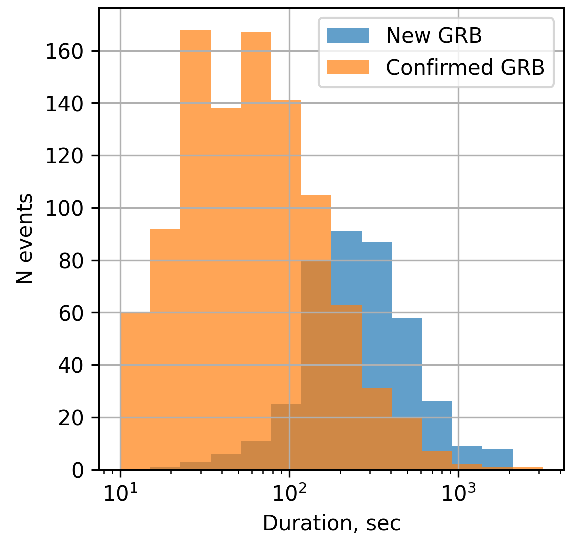}}
\caption{\small\rm The distributions of 403 potential ultra-long
  GRBs found in the SPI-ACS data by the developed method (blue)
  and 1018 GRBs identified with previously known events (orange)
  in duration $T_{1\sigma}$.}
\label{candidates_duration_distr}
\end{figure}

(3) We prepared two machine learning models: based on
classification with dimensionality reduction and on
clustering. Using them, we distinguished 551 events that most
likely belong to GRBs from 2429\,($=4364-1935$) potential
candidates for transients that did not match any events in any
of the catalogs being used. The precision of the first and second
models for the training sample is $\sim91\%$ and $\sim79\%$,
respectively.  However, the samples for training and testing the
model were not homogeneous and, hence, the precision estimates
could slightly differ from the actual values.
\begin{figure*}[p]
\center{~\includegraphics[width=0.78\linewidth]{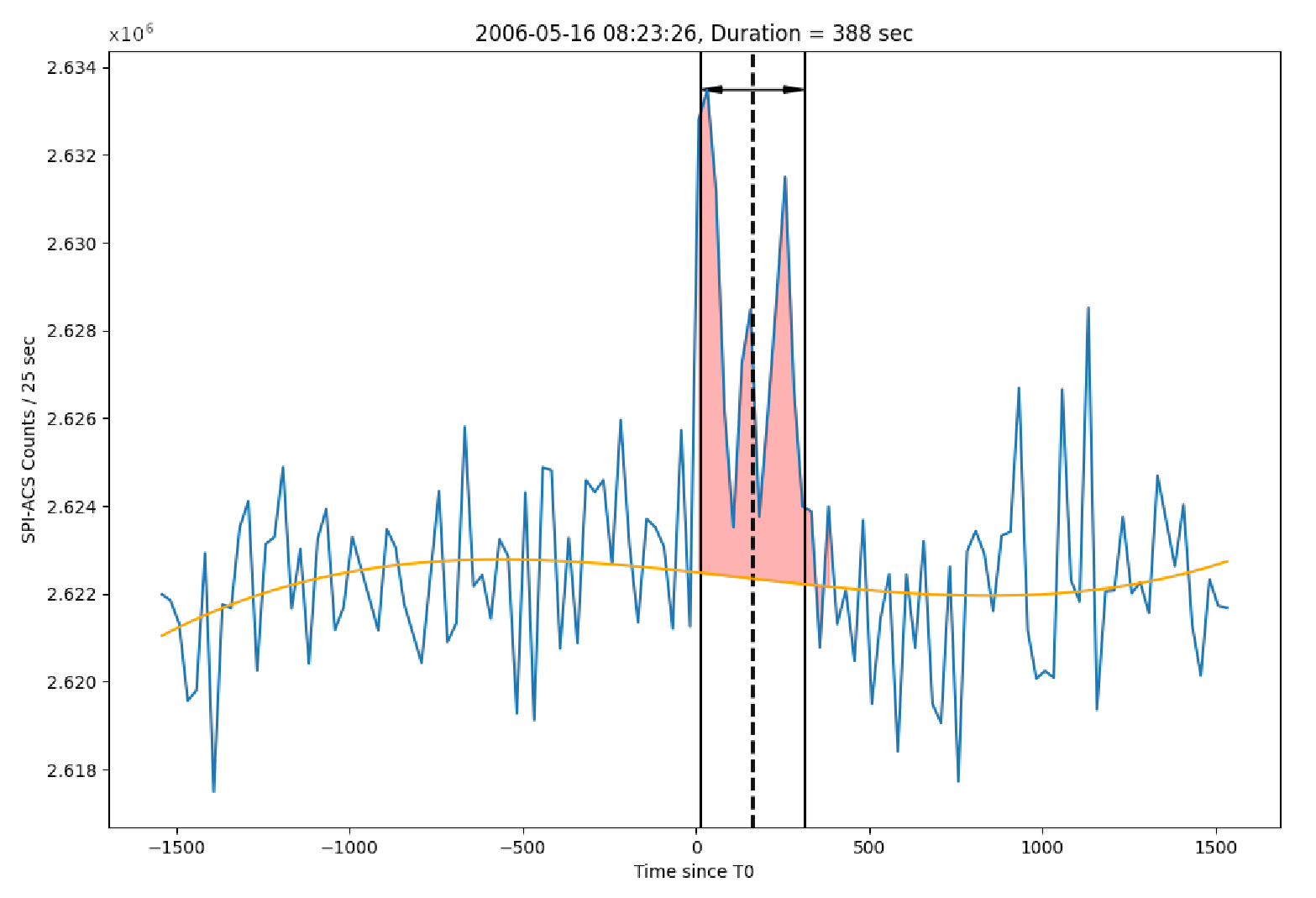}}
\caption{\small\rm The light curve of GRB~060516 --- one of the
  longest gamma-ray burst candidate from the SPI-ACS data. The
  red color highlights the event above the background. The
  orange color indicates the fit to the background by a
  third-degree polynomial. The black dashed line indicates the
  middle of the bin in which the event was found. The black
  solid lines indicate the left and right boundaries of this bin.}
\label{ultalong_1}
\vspace{4mm}
\center{\includegraphics[width=0.78\linewidth]{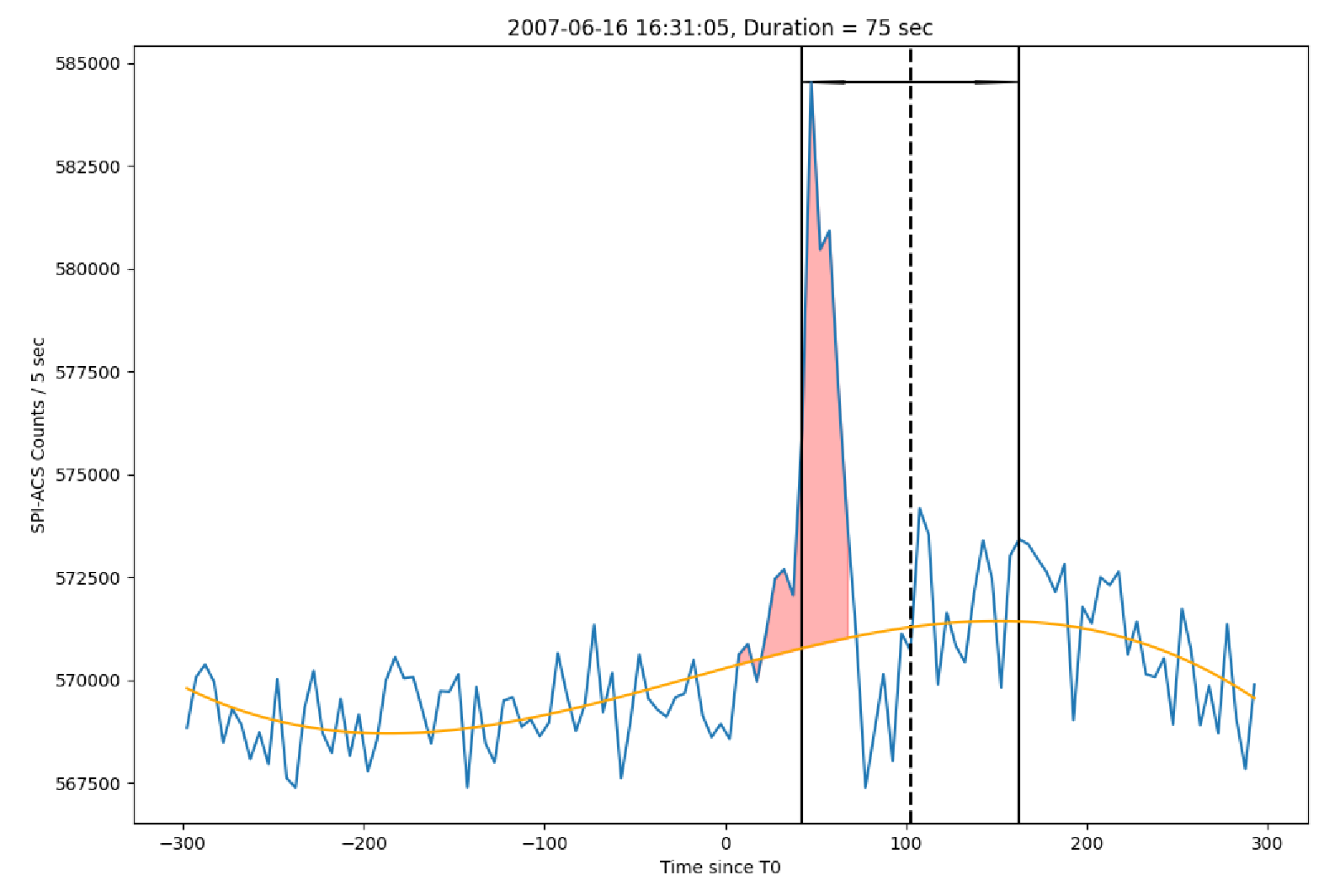}}
\caption{\small\rm Same as Fig.~\ref{ultalong_1}, but for the
  gamma-ray burst candidate GRB~070616.}\label{ultalong_2}
\end{figure*}

(4) We carried out additional studies to confirm the reality of
the revealed burst candidates. In particular, to search for and
confirm the events of interest to us, we reprocessed the
Konus-WIND archival data. Of the previously unknown transients
(551 events) detected by SPI-ACS and classified as probable
GRBs, we confirmed 17 events in the data of the Konus-WIND
instrument (that were not included in its GRB catalogs). Their
parameters are given in Table~\ref{real_candidates}. One of
these events was designated as a solar flare and, therefore, we
excluded it from further consideration.  Among the remaining 16
confirmed events, there are two candidates for the longest bursts
in our sample: May 16, 2006, 08:23:27 and June 16, 2007,
16:31:06; their duration from the ACS data is 388 and 75~s,
respectively. The light curves of these GRBs are presented in
Figs.~\ref{ultalong_1} and \ref{ultalong_2}.
\begin{table*}[t]
\caption{\small\rm The candidates for astrophysical transients
  found by the machine learning algorithms. Only the events
  confirmed by the Konus-WIND experiment are presented\ee.}
\label{real_candidates}
\hspace{0mm}
\footnotesize
\begin{tabular}{c|c|c|@{}r|l@{~}r@{}l|c|l@{~}l|c|l@{\,}}
\multicolumn{12}{c}{}\\ [-2mm]
 \hline
 $T_0,$&$T_{1\sigma},$&Time
 &\multicolumn{4}{c|}{Fluence\bb,}&\multicolumn{3}{c|}{Peak
 flux per 1 s\cc,}&Method\dd&Conf.\\ \cline{4-7}\cline{8-10}
&&&&\multicolumn{3}{c|}{}&&\multicolumn{2}{c|}{}&&\\ [-2mm]
UTC  & s&scale\aa, s&\ $10^3$ counts$\!\!$&\multicolumn{3}{c|}{$10^{-5}$
  erg cm$^{-2}$}&counts s$^{-1}$&\multicolumn{2}{c|}{$10^{-6}$
  erg cm$^{-2}$ s$^{-1}$} &   &  \\ \hline
&&&&&&&&&&&\\ [-3mm]
2003-04-27 09:57:20	&130&120&$44\pm	4$&$1.5$&$(+7.5,$&$-1.1)$&$1\,760\pm319$&$0.60$&$(+0.34,-0.05)$&HDBS &KW\\

2003-05-26 08:18:40	&120&300&$62\pm	3$&$2.1$&$(+10.3,$&$-1.6)$&$2\,302\pm320$&$0.79$&$(+0.42,-0.06)$&HDBS&KW\\

2003-06-13 17:14:45	&180&300&$65\pm	4$&$2.2$&$(+11.0,$&$-1.7)$&$1\,610\pm319$&$0.55$&$(+0.31,-0.04)$&HDBS&KW\\

2003-08-06 06:12:59	&388&300&$28\pm	6$&$1.0$&$(+5.5,$&$-0.8)$&$1\,223\pm318	$&$0.42$&$(+0.25,-0.03)$&HDBS&KW\\

2004-04-08 15:43:36	&43&120	&$24\pm 2$&$0.8$&$(+4.1,$&$-0.6)$&$1\,788\pm319 $&$0.61$&$(+0.34,-0.05)$&HDBS&KW\\
&&&&&&&&&&UMAP&\\
2004-07-09 00:59:43	&15&120	&$26\pm
1$&$0.9$&$(+4.3,$&$-0.7)$&$6\,419\pm326
$&$2.20$&$(+1.07,-0.16)$&HDBS&KW\\
&&&&&&&&&&UMAP&	\\
2004-12-24 17:38:58	&389&300&$57\pm	6$&$1.9$&$(+10.1,$&$-1.5)$&$1\,304\pm318$&$0.45$&$(+0.27,-0.04)$&HDBS&	\\

2005-01-05 16:59:28
&389&300&$102\pm6$&$3.5$&$(+17.2,$&$-2.6)$&$2\,611\pm320$&$0.90$&$(+0.47,-0.07)$&HDBS&IBAS,\\
&&&&&&&&&&&KW\\

2005-07-03 19:58:16	&75&120	&$49\pm	3$&$1.7$&$(+8.1,$&$-1.2)$&$2\,957\pm321$&$1.01$&$(+0.52,-0.08)$&HDBS&KW\\
&&&&&&&&&&UMAP&	\\
2006-02-25 15:16:28	&56&300	&$27\pm	2$&$0.9$&$(+4.7,$&$-0.7)$&$1\,533\pm319$&$0.53$&$(+0.30,-0.04)$&HDBS&KW\\

2006-05-16 08:23:27	&389&300&$63\pm	6$&$2.2$&$(+11.1,$&$-1.6)$&$1\,494\pm319$&$0.51$&$(+0.29,-0.04)$&HDBS&KW\\

2007-06-16 16:31:06	&75&120	&$70\pm	3$&$2.4$&$(+11.4,$&$-1.8)$&$3\,413\pm322$&$1.17$&$(+0.60,-0.09)$&HDBS&KW\\
&&&&&&&&&&UMAP&	\\
2008-02-26 16:13:57	&478&300&$43\pm	7$&$1.5$&$(+8.1,$&$-1.1)$&$1\,482\pm319	$&$0.51$&$(+0.29,-0.04)$&HDBS&KW\\

2008-04-13 21:19:49	&170&120&$30\pm	4$&$1.0$&$(+5.5,$&$-0.8)$&$2\,206\pm320	$&$0.76$&$(+0.41,-0.06)$&HDBS&KW\\
&&&&&&&&&&UMAP&	\\
2009-03-23 09:54:57	&85&120	&$33\pm	3$&$1.1$&$(+5.7,$&$-0.9)$&$1\,799\pm319	$&$0.62$&$(+0.34,-0.05)$&HDBS&KW\\
&&&&&&&&&&UMAP&	\\
2011-12-31 13:41:49	&20&120	&$15\pm	1$&$0.5$&$(+2.7,$&$-0.4)$&$4\,452\pm323	$&$1.53$&$(+0.76,-0.11)$&UMAP&IBAS,\\
&&&&&&&&&&&KW\\

2015-11-20 00:09:05	&85&120	&$19\pm 3$&$0.7$&$(+3.5,$&$-0.5)$&$1\,164\pm318	$&$0.40$&$(+0.24,-0.03)$&HDBS&KW\\
&&&&&&&&&&UMAP&	\\ \hline

\multicolumn{12}{l}{}\\ [-2mm]
\multicolumn{12}{l}{\aa\ The time scale on which the transient was detected.}\\ 
\multicolumn{12}{l}{\bb\ The fluence in the event time
  $T_{1\sigma}$ from the SPI-ACS data. The flux estimation 
  technique is described in Minaev et al. (2023).}\\
\multicolumn{12}{l}{\cc\ The event peak flux on a time scale of
  1~s from the SPI-ACS data.}\\  
\multicolumn{12}{l}{\dd\ The machine event classification method:
  HDBSCAN clustering and UMAP dimensionality reduction.}\\ 
\multicolumn{12}{l}{\ee\ \,A full version of the table is
  accessible in electronic form at grb.rssi.ru/INTEGRAL/GRB\_ACS\_candidates.txt.}\\ 
\end{tabular}
\end{table*}

(5) The non-detection of the remaining 534 events in the
Konus-WIND data by no means implies that they are all background
or solar ones. Konus-WIND might not detect them because of its
insufficient sensitivity compared to the sensitivity of the
SPI-ACS detector. To check this, let us compare the Konus-WIND
(Kozlova et al. 2019) and SPI-ACS event detection thresholds (we
use the calibration from Minaev and Pozanenko (2023) for the
conversion to energy units). For simplicity, we will compare the
maximum Konus-WIND detector threshold and the minimum SPI-ACS
conversion coefficient and will find that, indeed, Konus-WIND
could not detect 403 ($75\%$) of the events under
discussion. Since the remaining events, be they GRBs, must have
been detected by this detector at a $\sim4\sigma$ confidence
level, we conclude that they all (131 events) were caused in the
SPI-ACS data by local geophysical factors.

(6) The fraction of actual GRBs among the 403 mentioned
candidates can be estimated independently. For this purpose, it
is necessary to calculate the fraction of background events in
the test sample. This can be done if the model precision is known
($91\%$). Let $\alpha$ be the number of GRBs in the test
sample. Then,
\begin{equation}
    (2429-\alpha)\times 0.09 + \alpha\times 0.91 = 551,
\end{equation}
and, accordingly,
\begin{equation}
    \alpha = \frac{551 - 2429\times 0.09}{0.91 - 0.09} = 405.
\end{equation}
Thus, $405\times0.91=369$ ($67\%$) of the 551 events were
actually obtained from the sample of GRBs. This implies that
among the 403 events, up to 270 can be actual GRBs.   

\section{DISCUSSION AND CONCLUSIONS}
\noindent
We carried out a blind search for long transient events on time
scales of 120, 300, and 1000~s based on the SPI-ACS data
spanning $\sim20$ years of INTEGRAL operation. We classified the
transient events found with the help of machine learning using
information from other INTEGRAL detectors (ISGRI and
IREM). Owing to the use of machine learning models, out of all
4364 candidates for transient events found, we managed to
independently classify 1018 already known GRBs from other
catalogs (their duration distribution is presented in
Fig.~\ref{candidates_duration_distr}) and 2429 previously
undetected GRB candidates. For 551 events identified by the
models as candidates for astrophysical events we carried out a
search in the Konus-WIND data and found significant synchronous
excesses in the light curves in one of the two Konus-WIND
detectors for 17 events. After a detailed analysis, one of these
17 simultaneously recorded events was identified with a solar
flare. Thus, for 16 events we conformed their astrophysical
nature as cosmic GRBs; these events have not been recorded
previously by space gamma-ray experiments.

We showed that after the elimination of the probable background
and solar flares from the remaining 534 events, the sample must
still contain 403 previously unknown candidates for
astrophysical transients that have not been confirmed by any
catalogs or data from the Konus-WIND experiment, up to 270
events of which may turn out to be real cosmic GRBs. The
statistics of the results of the performed analysis are
summarized in Table~\ref{ident}. All these candidates were found
on search time scales of more than 120~s. Among the classified
1018 already known GRBs found independently in this paper using
the same algorithms, five events have a duration of more than
900~s. Thus, ultra-long GRBs are actually detected in the
SPI-ACS data at energies $>80$ keV, both among the already known
1018 GRBs and among the 403 GRB candidates.

\begin{table}
\caption{\small\rm Summary statistics on the results of our search
and the classification of transients.} 
\label{ident}

\centering
\begin{tabular}{l|r}
  \multicolumn{2}{c}{}\\ [-2mm]
  \hline
  & \\ [-3mm]
\small
  Total number of excesses above the background & 4364 \\
 Already known & 1935\\
 \hspace{1cm}Known GRBs among them&1018\\
 Candidates & 2429\\
 Identified by the machine learning models & 551 \\
 Coincidences in time with Konus-WIND & 17 \\
 \hspace{1cm}GRBs among them & 16\\
 Candidates without confirmation  & 403 \\ 
 \hspace{1cm}GRBs expected among them& $\leq$  270\\ \hline
\end{tabular}
\end{table}

Only 16 of the previously unknown events found in the SPI-ACS
data were confirmed in the Konus-WIND data during our additional
analysis. This allows these 16 events to be classified with a high
probability as GRBs, but only four of them have a duration of
more than 350~s. The significant difference between the number of
SPI-ACS candidates (403) and the number of matches with Konus-WIND
(16 GRBs and one solar flare) is probably related to the
selection effects due to the lower detection energy threshold in
the Konus-WIND experiment (20 keV) than the SPI-ACS detection
threshold (80 keV). 

\section*{ACKNOWLEDGMENTS}
\label{Acknowledgments}
\noindent 
G.Yu. Mozgunov and S.A. Grebenev are grateful to the ``BASIS''
Foundation for the Theoretical Physics and Mathematics
Advancement, grant no. 22-1-1-57-1 of the ``Leading Scientist''
Program (theoretical physics) for supporting the development of
the light-curve processing algorithms using machine learning
methods.

A.S. Pozanenko and P.Yu. Minaev are grateful to the Russian
Science Foundation (grant no. 23-12-00198) for supporting the
work with regard to the analysis of the observations of GRBs in
the GBM/Fermi and BAT/Swift experiments and Kevin Hurley's
Masterlist. The work of A.G. Demin, A.V. Ridnaia, D.S. Svinkin,
and D.D. Frederiks was performed within the State Assignment
Theme of the Ioffe Physical-Technical Institute
(FFUG-2024-0002).

\section*{CONFLICT OF INTEREST}
\noindent
The authors of this work declare that they have no conflicts of
interest.

\begin{figure*}[!t]
\center{\includegraphics[width=0.62\linewidth]{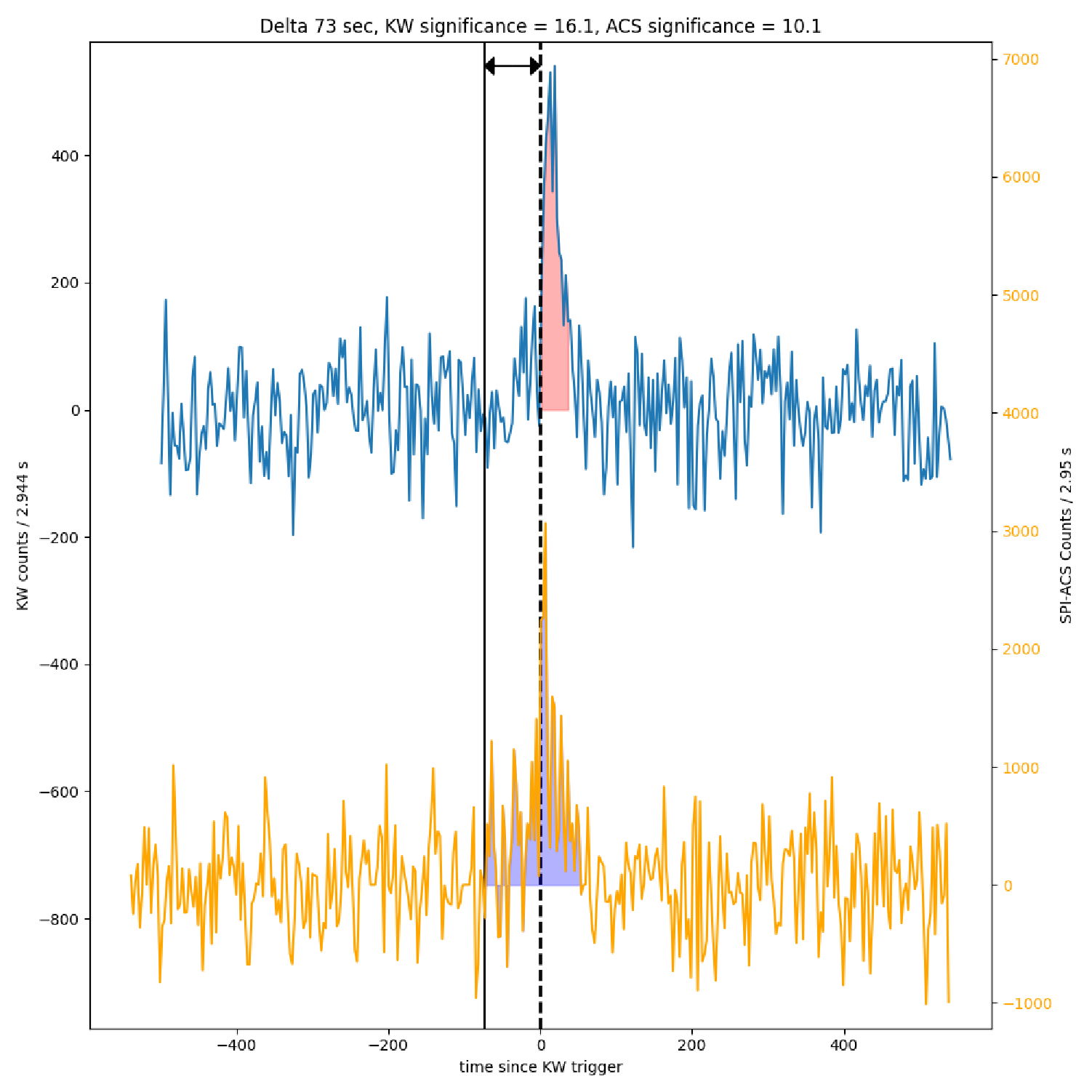}}
\caption{\rm Comparison of the light curves for the gamma-ray burst
  candidate GRB~030427 from the Konus-WIND (the blue line, the Y
  axis on the left) and SPI-ACS (the yellow line, the Y axis on
  the right) data. The black dashed line indicates the
  Konus-WIND event trigger time; the black solid line indicates
  the left boundary of the bin in which the burst was found when
  analyzing the SPI-ACS data.}\label{firstgrb}
\end{figure*}

\vspace{2mm}

\begin{flushright}
{\sl Translated by V. Astakhov\/}
\end{flushright}

\vspace{4mm}

\begin{flushright}
{\bf APPENDIX}
\end{flushright}

\vspace{2mm}

\noindent 
To illustrate the results presented above, Figs.~9--23 show the
light curves of the gamma-ray burst candidates found from the
SPI-ACS data using the machine learning algorithms in comparison
with the light curves recorded at the same time by the
Konus-WIND detector. The parameters of most of these events are
given in Table~\ref{real_candidates}.

\clearpage
\begin{figure*}[t]
\center{\includegraphics[width=0.62\linewidth]{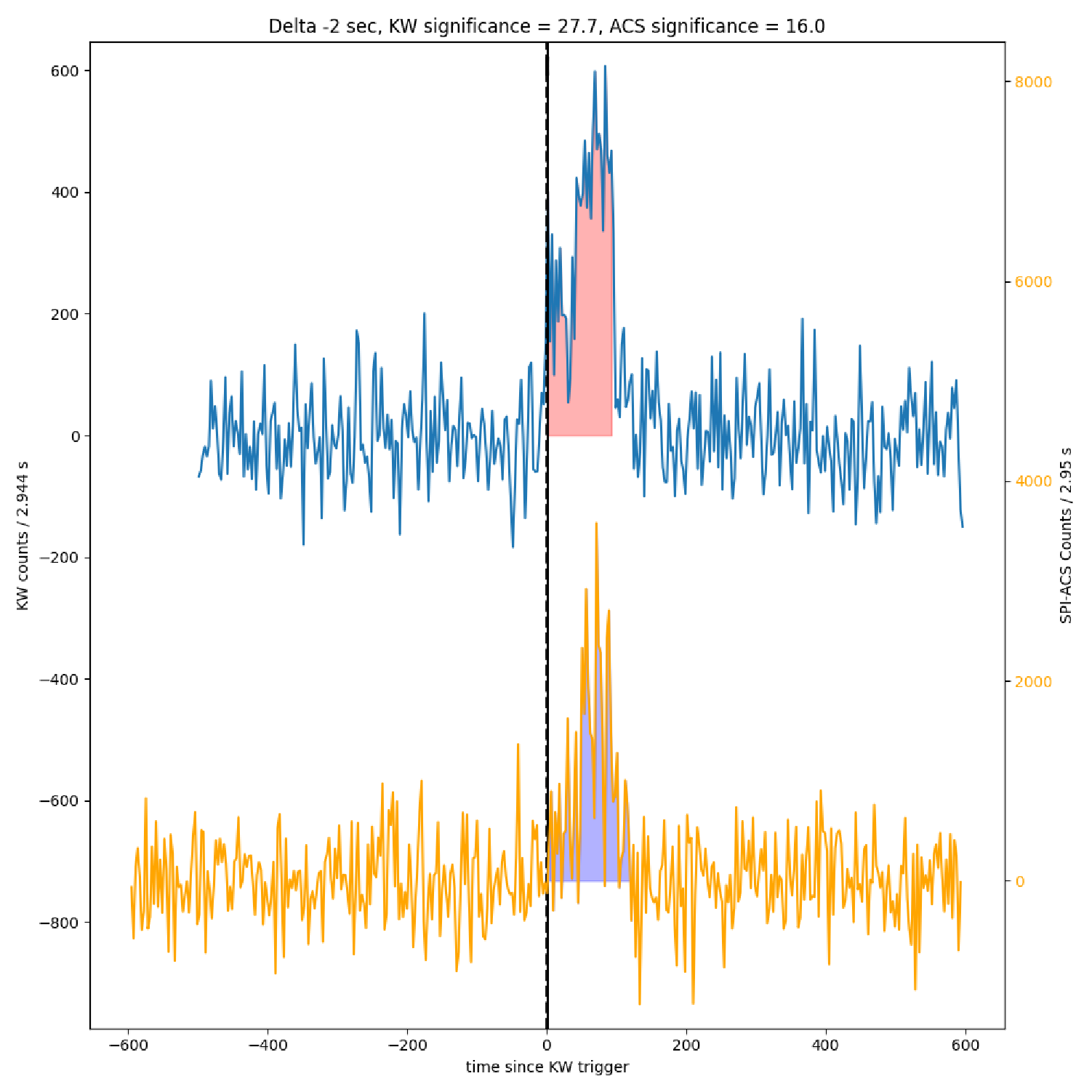}}
\caption{\rm Same as Fig.~\ref{firstgrb}, but for the gamma-ray
  burst candidate GRB~030526.} 
\end{figure*}
\begin{figure*}[h]
\center{\includegraphics[width=0.62\linewidth]{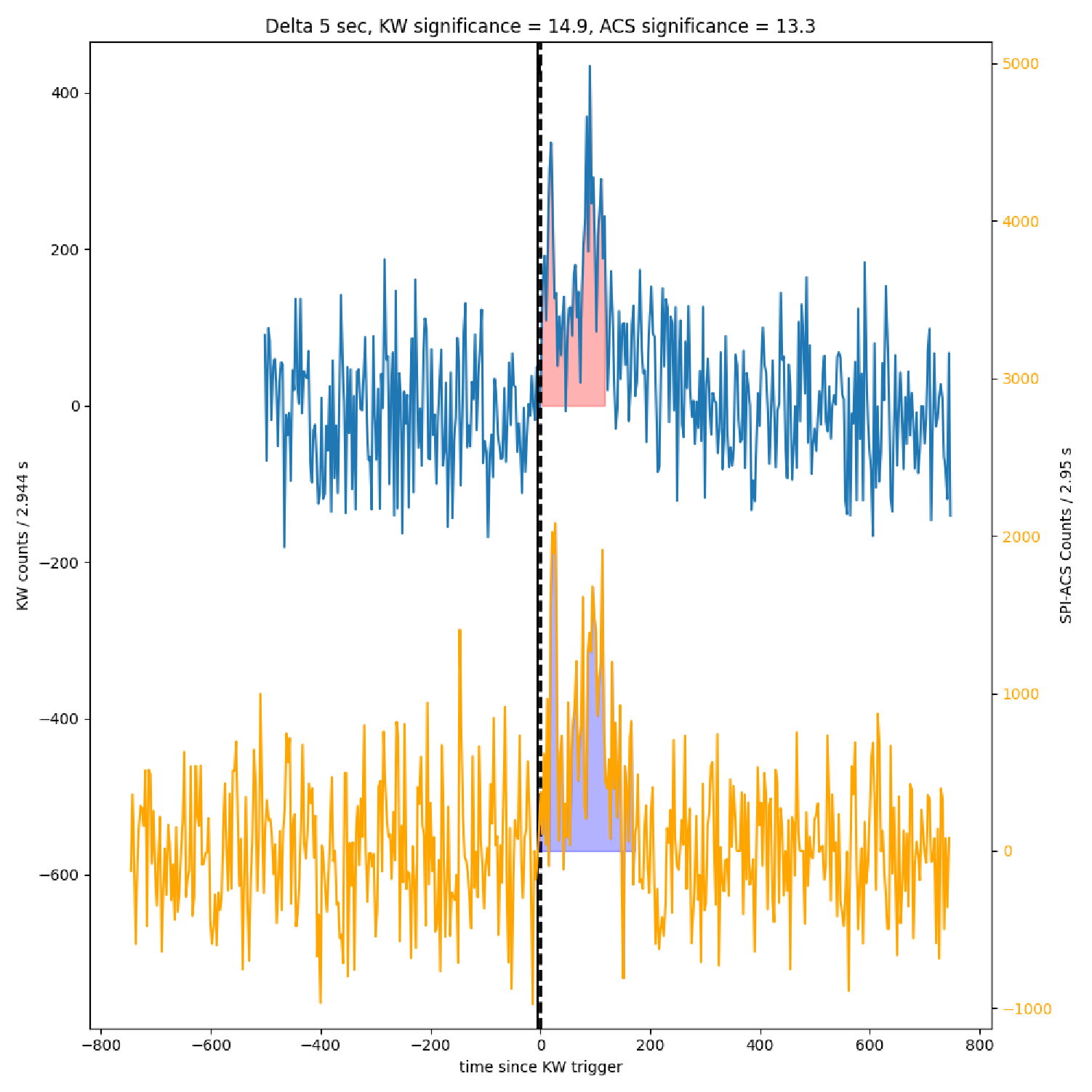}}
\caption{\rm Same as Fig.~\ref{firstgrb}, but for the gamma-ray
  burst candidate GRB~030613.} 
\end{figure*}
\clearpage
\begin{figure*}[t]
\center{\includegraphics[width=0.62\linewidth]{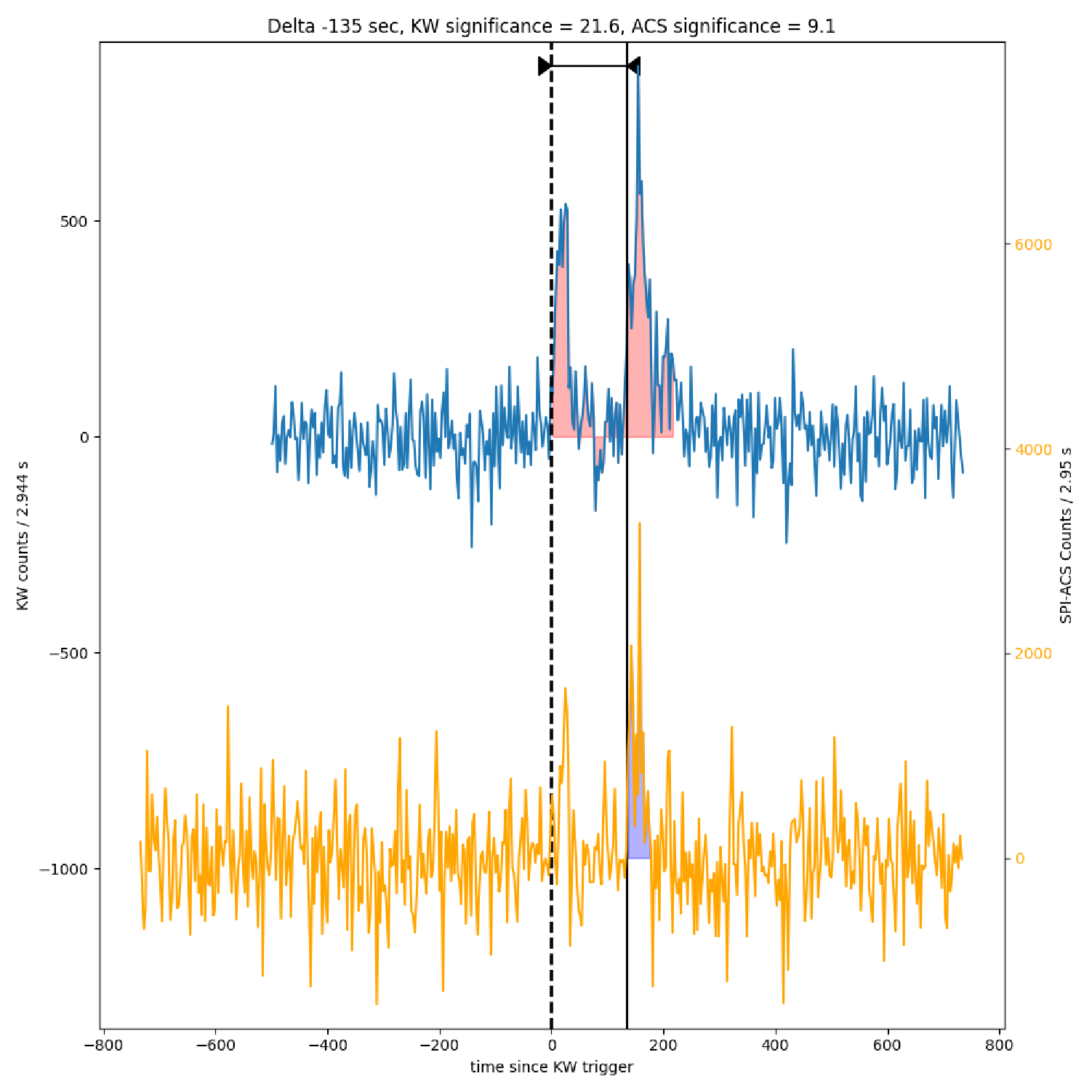}}
\caption{\rm Same as Fig.~\ref{firstgrb}, but for the gamma-ray
  burst candidate GRB~040408.} 
\end{figure*}
\begin{figure*}[h]
\center{\includegraphics[width=0.62\linewidth]{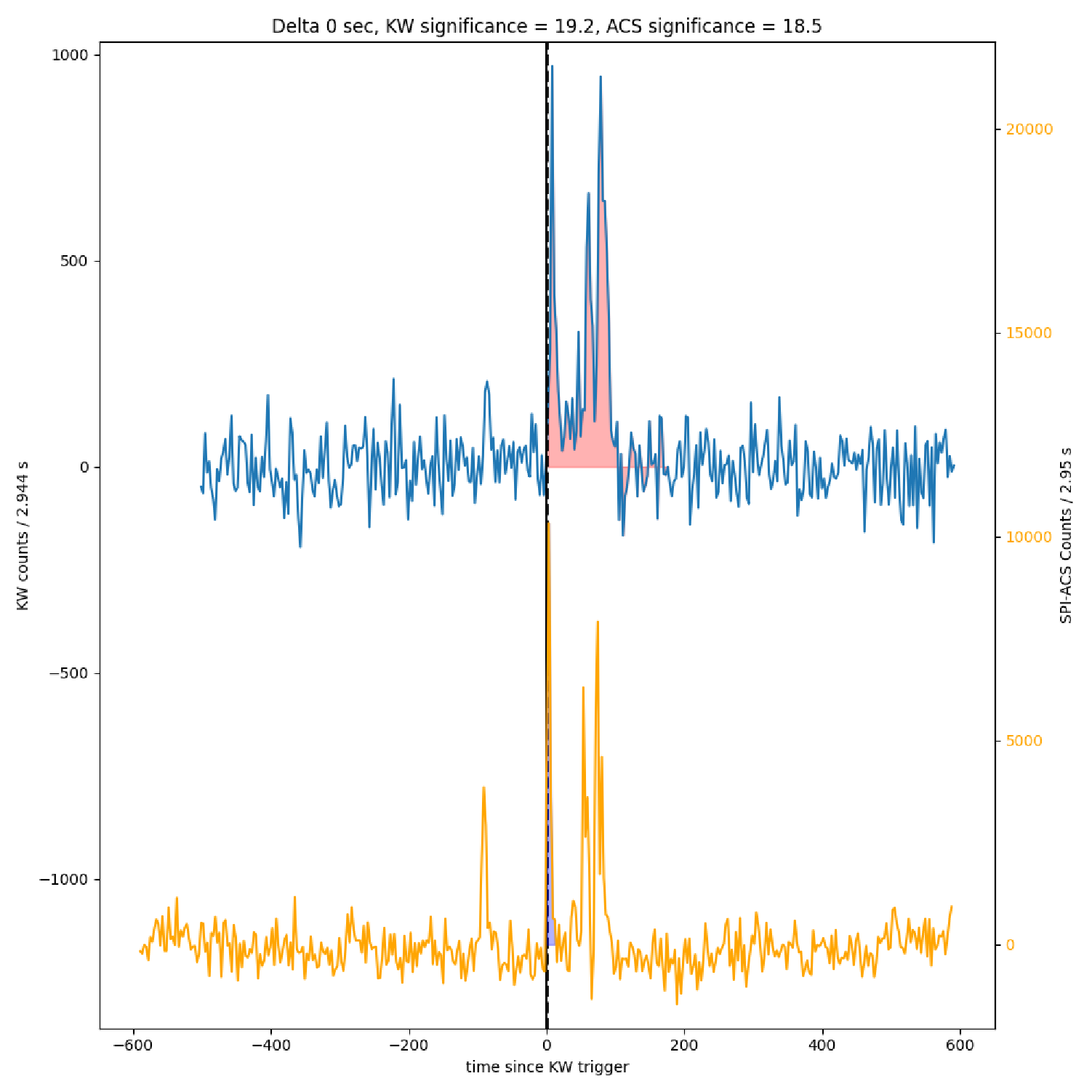}}
\caption{\rm Same as Fig.~\ref{firstgrb}, but for the gamma-ray
  burst candidate GRB~040709.} 
\end{figure*}
\clearpage
\begin{figure*}[t]
\center{\includegraphics[width=0.62\linewidth]{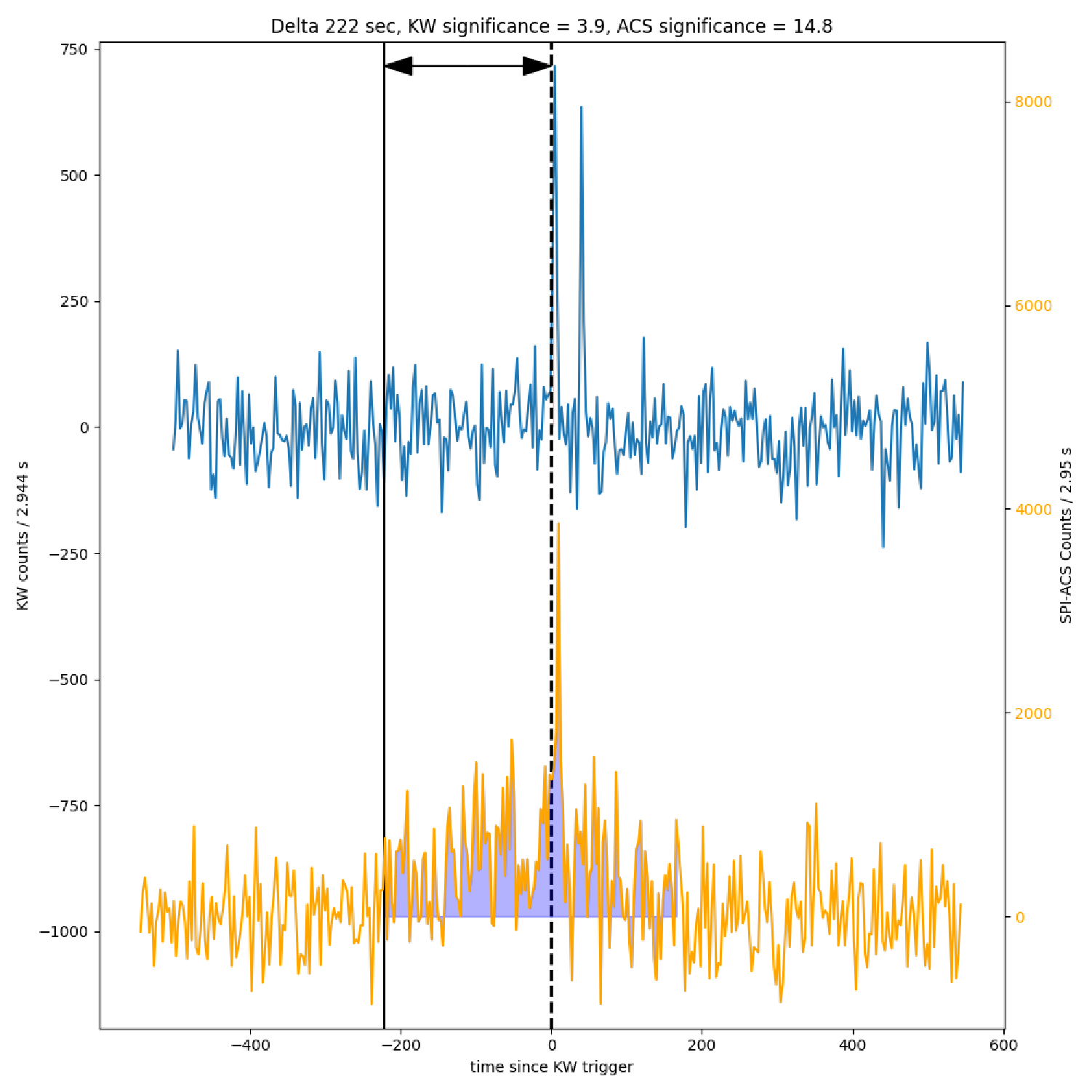}}
\caption{\rm Same as Fig.~\ref{firstgrb}, but for the gamma-ray
  burst candidate GRB~050105.} 
\end{figure*}
\begin{figure*}[h]
\center{\includegraphics[width=0.62\linewidth]{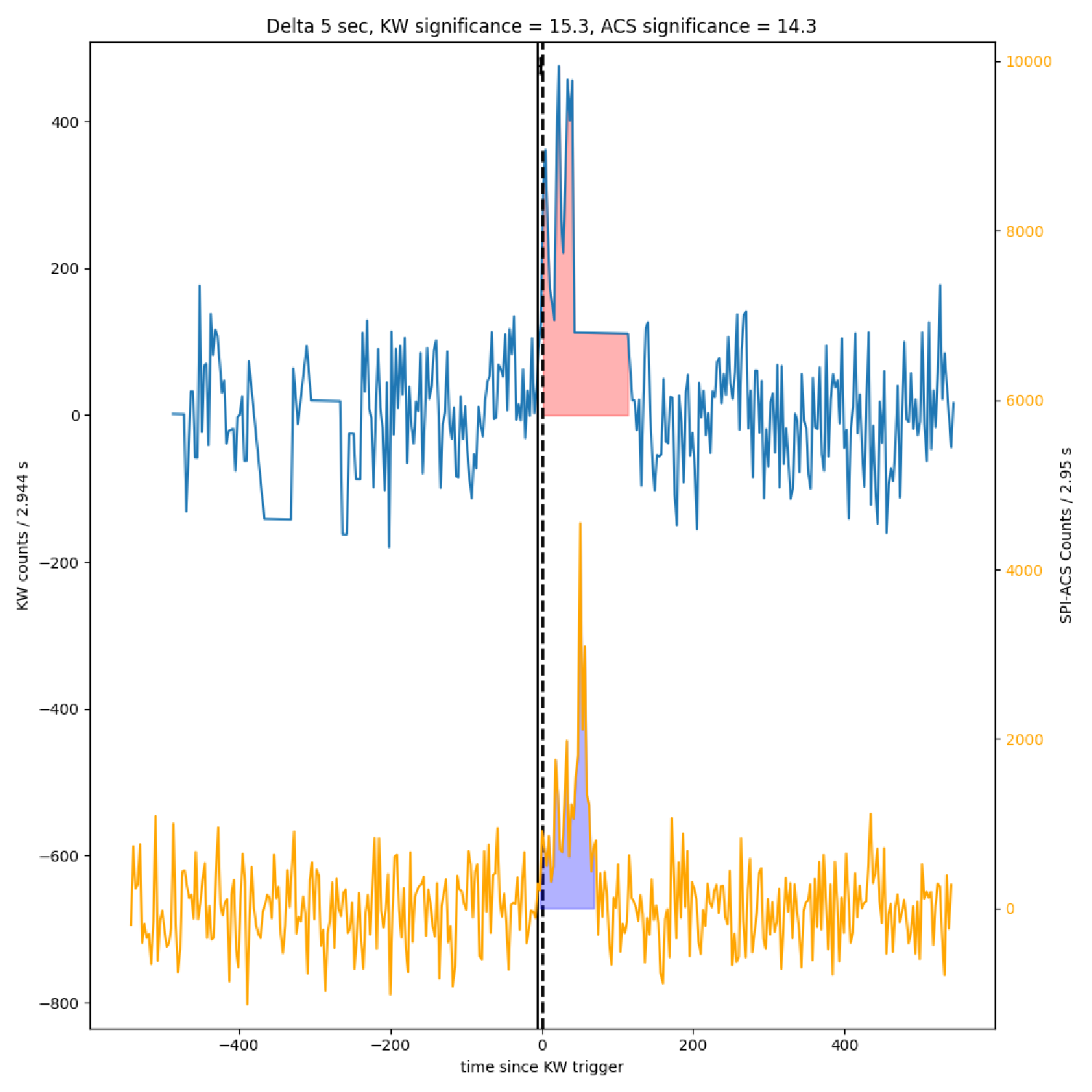}}
\caption{\rm Same as Fig.~\ref{firstgrb}, but for the gamma-ray
  burst candidate GRB~050703.} 
\end{figure*}
\clearpage
\begin{figure*}[t]
\center{\includegraphics[width=0.62\linewidth]{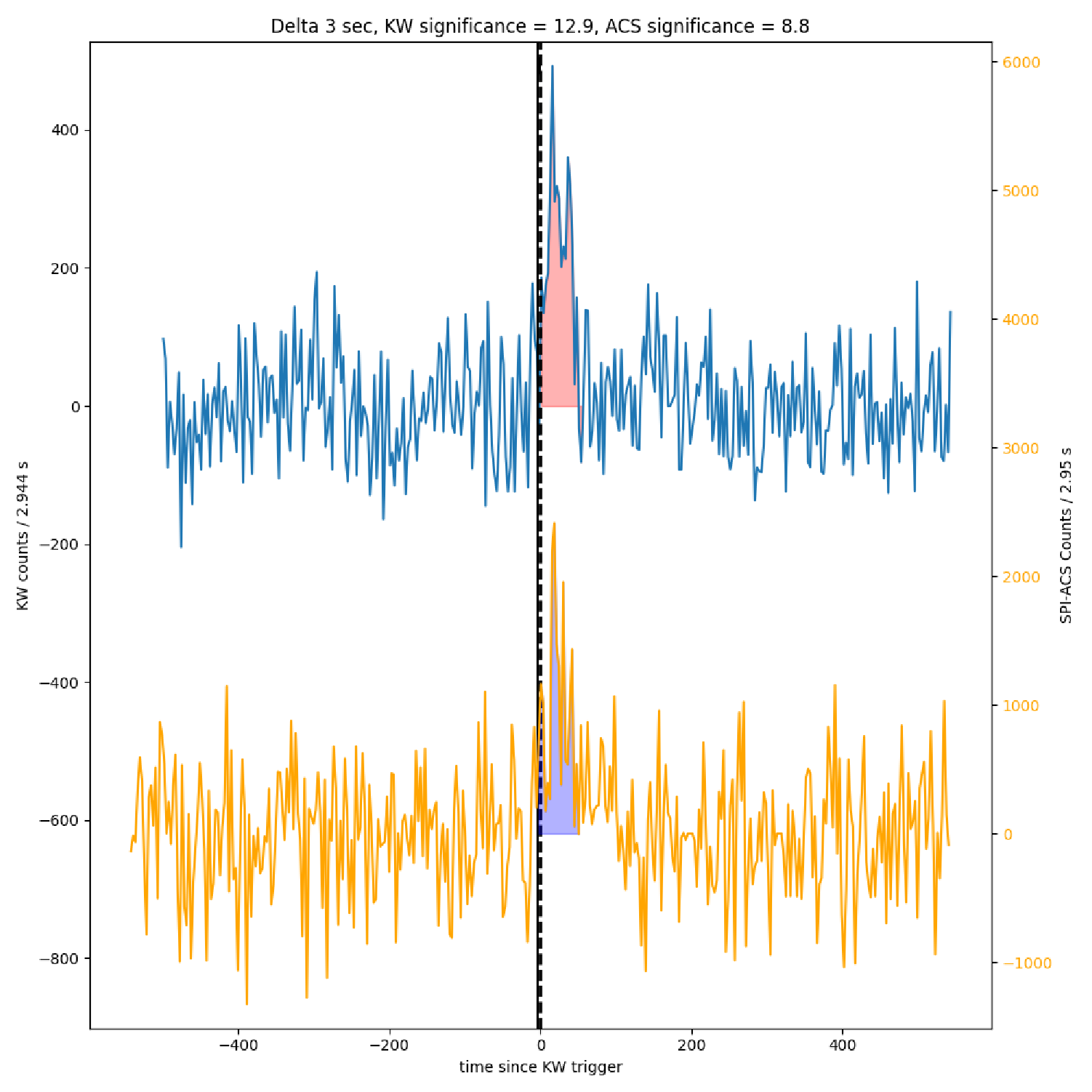}}
\caption{\rm Same as Fig.~\ref{firstgrb}, but for the gamma-ray
  burst candidate GRB~060225.} 
\end{figure*}
\begin{figure*}[h]
\center{\includegraphics[width=0.62\linewidth]{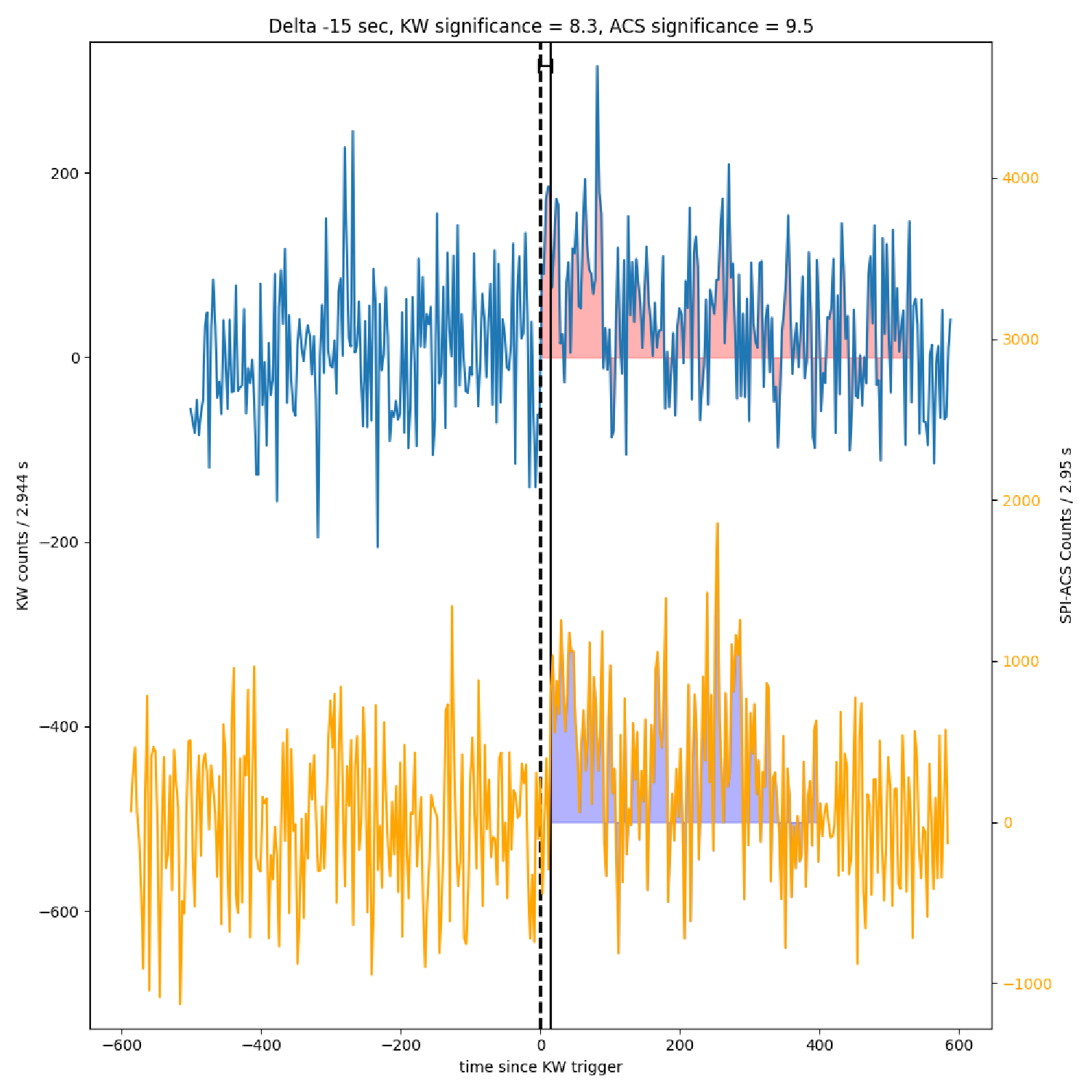}}
\caption{\rm Same as Fig.~\ref{firstgrb}, but for the gamma-ray
  burst candidate GRB~060516.} 
\end{figure*}
\clearpage
\begin{figure*}[t]
\center{\includegraphics[width=0.62\linewidth]{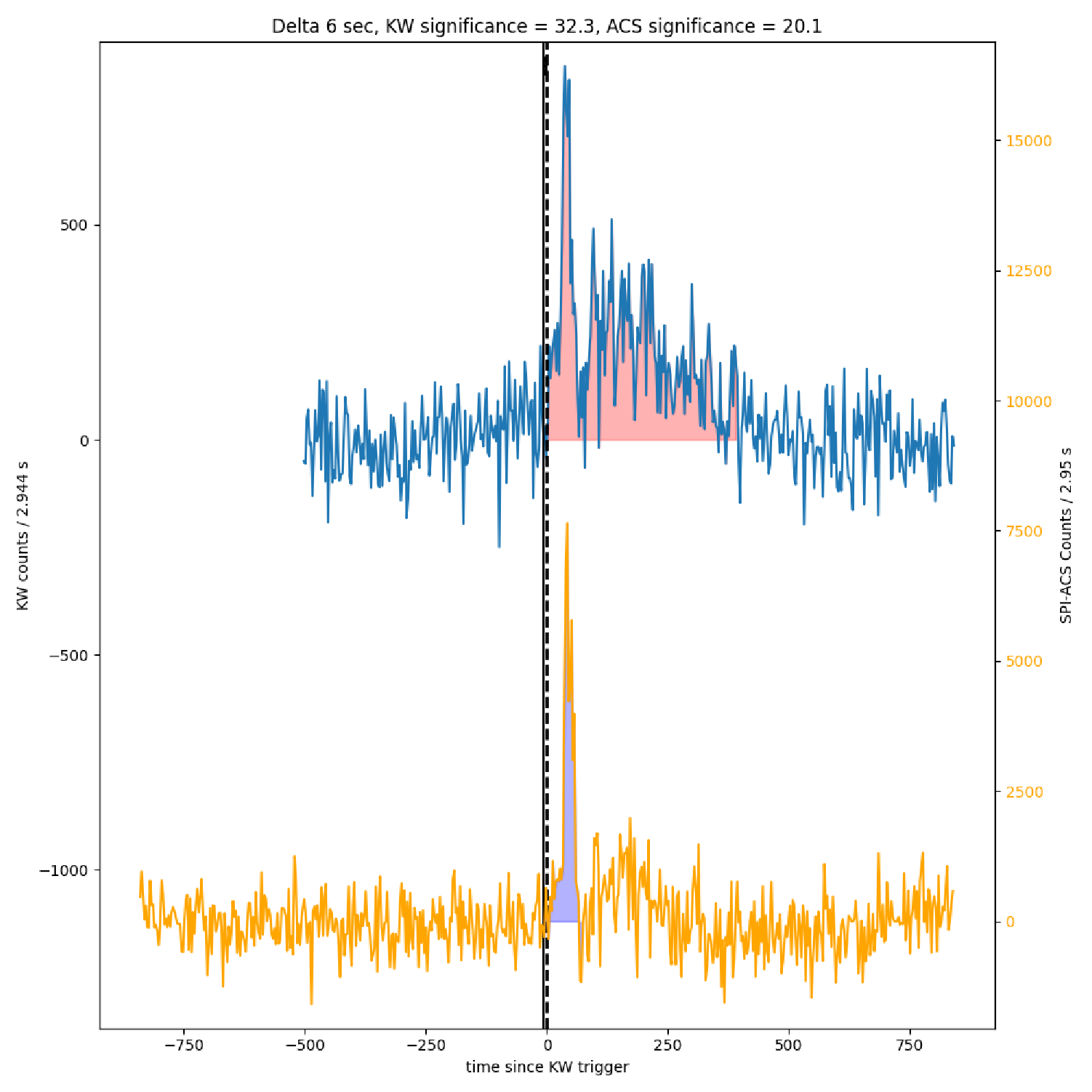}}
\caption{\rm Same as Fig.~\ref{firstgrb}, but for the gamma-ray
  burst candidate GRB~070616.} 
\end{figure*}
\begin{figure*}[h]
\center{\includegraphics[width=0.62\linewidth]{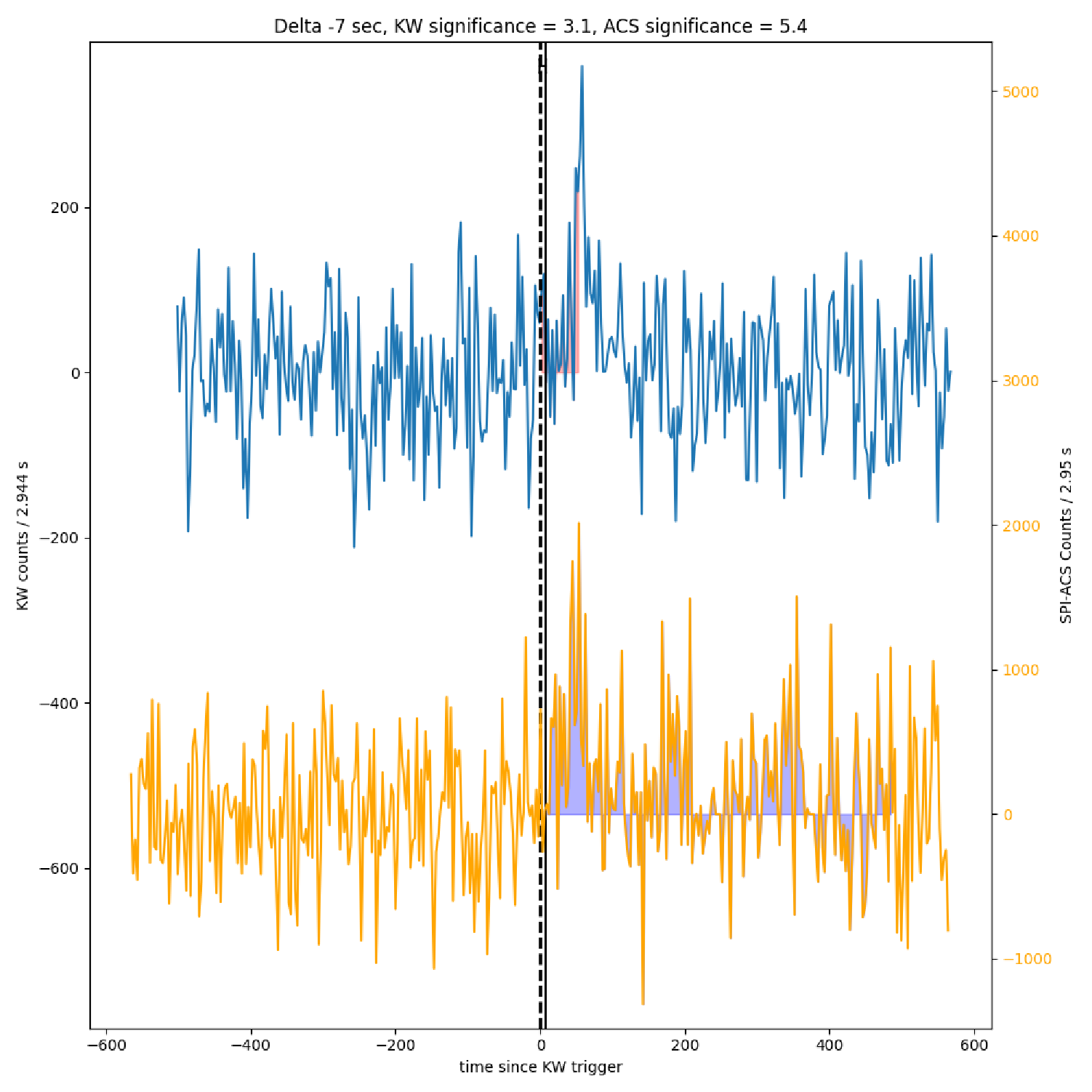}}
\caption{\rm Same as Fig.~\ref{firstgrb}, but for the gamma-ray
  burst candidate GRB~080226.}
\end{figure*}
\clearpage
\begin{figure*}[t]
\center{\includegraphics[width=0.62\linewidth]{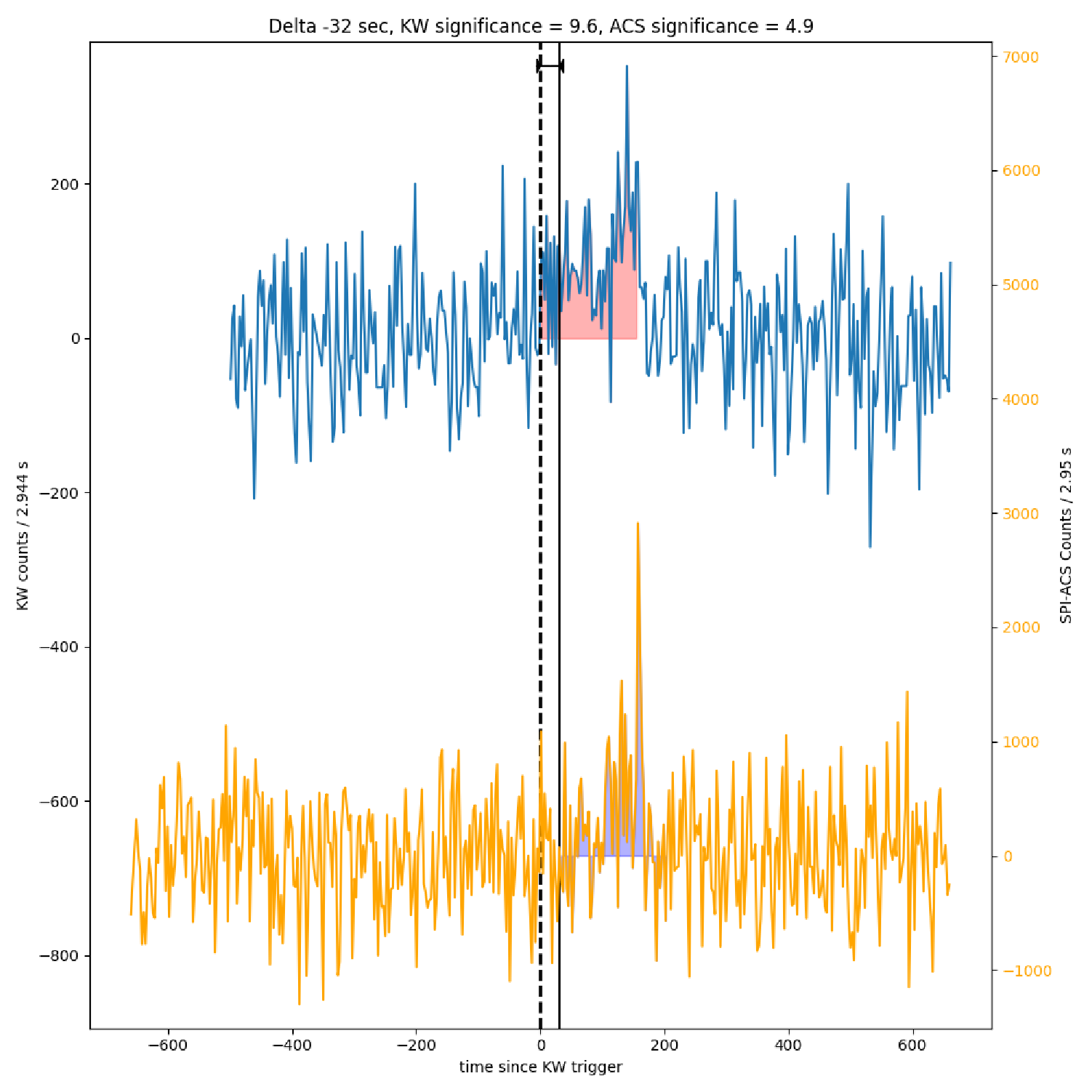}}
\caption{\rm Same as Fig.~\ref{firstgrb}, but for the gamma-ray
  burst candidate GRB~080413.}
\end{figure*}
\begin{figure*}[h]
\center{\includegraphics[width=0.62\linewidth]{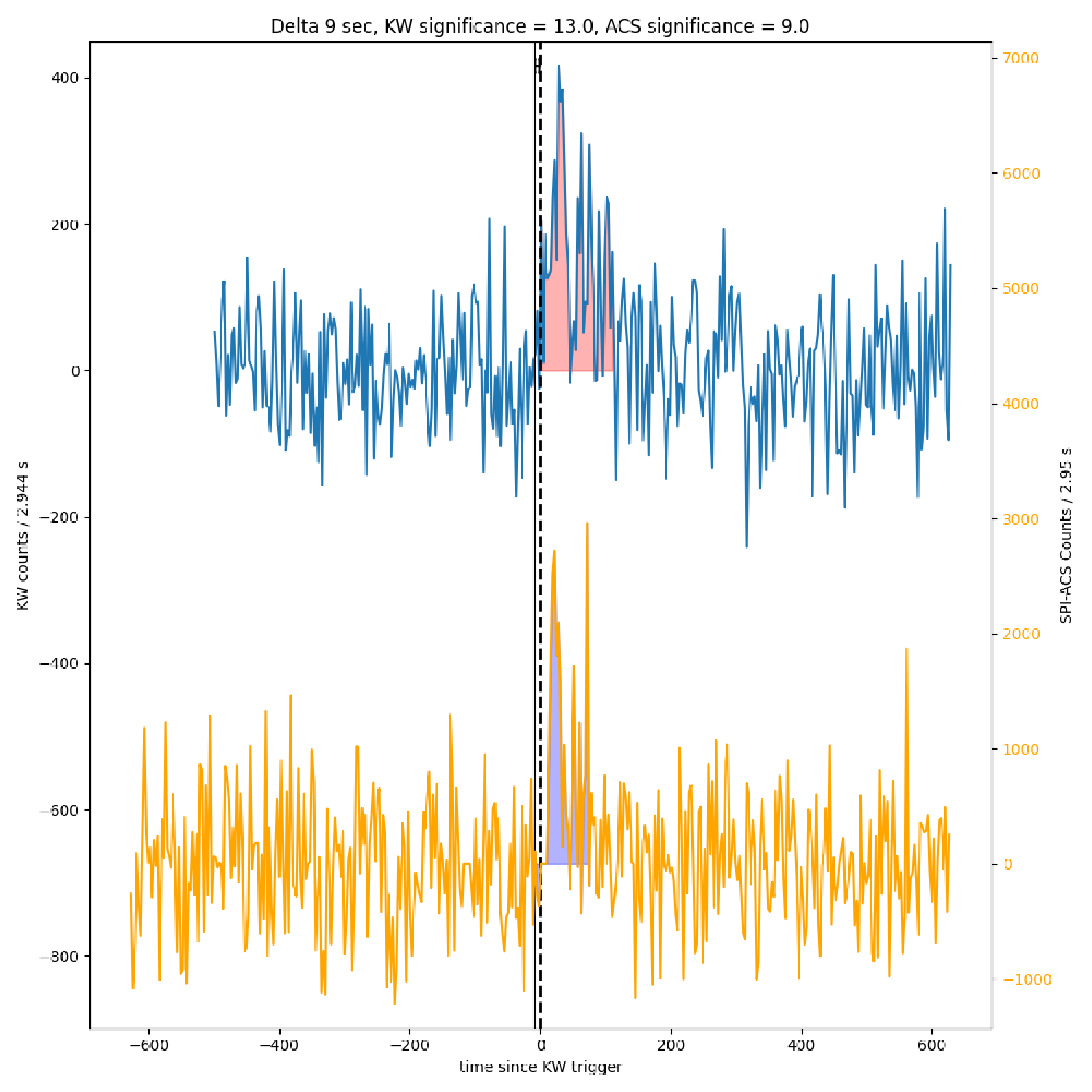}}
\caption{\rm Same as Fig.~\ref{firstgrb}, but for the gamma-ray
  burst candidate GRB~090323.}
\end{figure*}
\clearpage
\begin{figure*}[t]
\center{\includegraphics[width=0.62\linewidth]{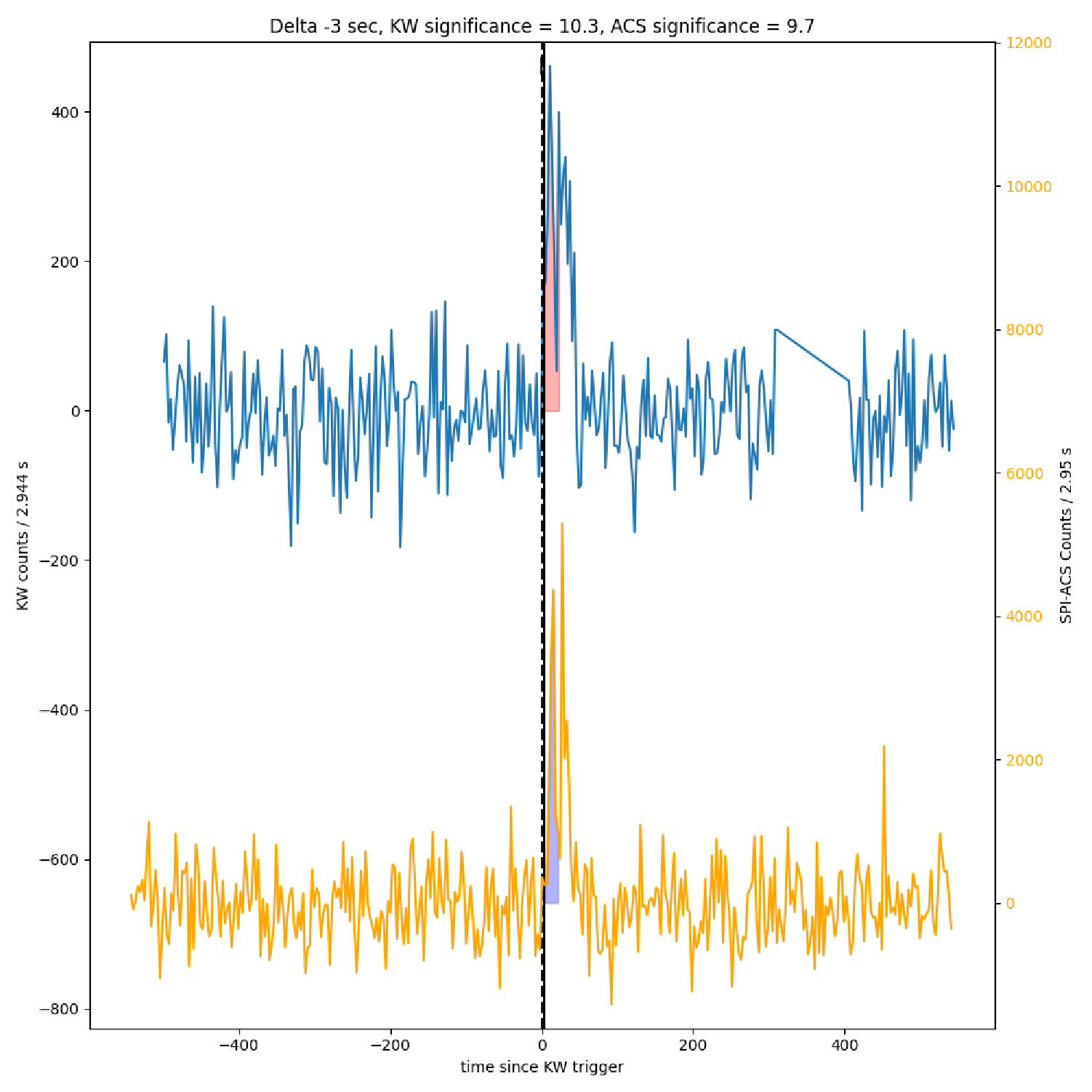}}
\caption{\rm Same as Fig.~\ref{firstgrb}, but for the gamma-ray
  burst candidate GRB~111231.}
\end{figure*}
\begin{figure*}[h]
\center{\includegraphics[width=0.62\linewidth]{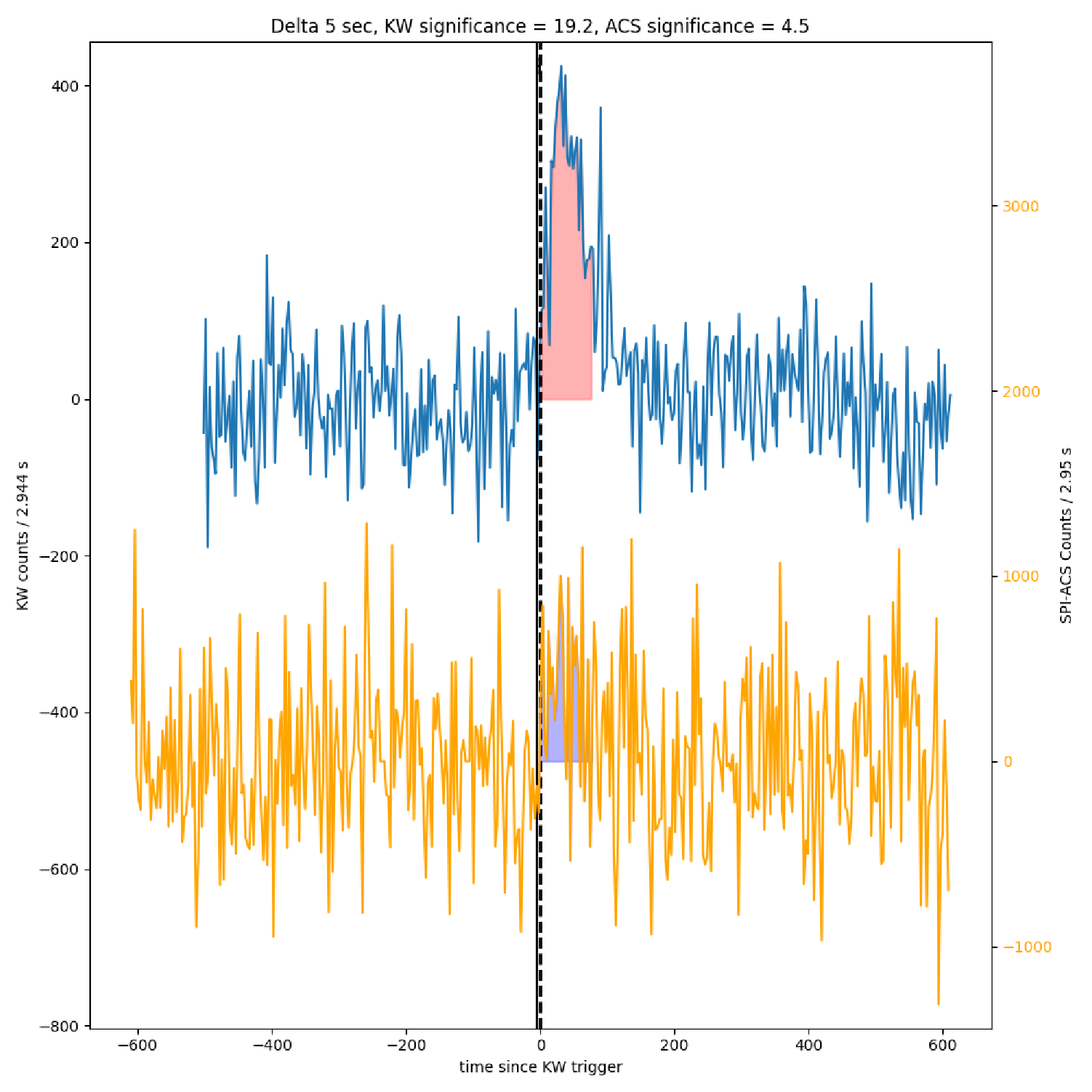}}
\caption{\rm Same as Fig.~\ref{firstgrb}, but for the gamma-ray
  burst candidate GRB~151120.}
\end{figure*}
\end{document}